\documentclass[usenatbib,usegraphicx]{mn2e}

\usepackage{upgreek}

\newcommand{\ion}[2]{\mbox{#1\,\textsc{#2}}}

\title[Gas and dark matter in the Sculptor group: NGC~55]{Gas and dark matter in the Sculptor group: NGC~55}
\author[T.~Westmeier, B.~S.~Koribalski, and R.~Braun]{T.~Westmeier$^{1,2}$\thanks{E-mail:
tobias.westmeier@uwa.edu.au}, B.~S.~Koribalski$^{2}$, and R.~Braun$^{2}$\\
$^{1}$ICRAR, M468, The University of Western Australia, 35~Stirling Highway, Crawley~WA~6009, Australia\\
$^{2}$Australia Telescope National Facility, CSIRO Astronomy and Space Science, PO~Box~76, Epping~NSW~1710, Australia}

\begin{document}
  \date{Accepted 1988 December 15. Received 1988 December 14; in original form 1988 October 11}
  
  \pagerange{\pageref{firstpage}--\pageref{lastpage}} \pubyear{2002}
  
  \maketitle
  
  \label{firstpage}

  \begin{abstract}
    We present new, sensitive \ion{H}{i}~observations of the Sculptor group galaxy NGC~55 with the Australia Telescope Compact Array. We achieve a $5 \sigma$ \ion{H}{i}~column density sensitivity of $10^{19}~\mathrm{cm}^{-2}$ over a spectral channel width of $8~\mathrm{km \, s}^{-1}$ for emission filling the $158\arcsec{} \! \times 84\arcsec{}$ synthesised beam. Our observations reveal for the first time the full extent of the \ion{H}{i}~disc of NGC~55 at this sensitivity and at a moderately high spatial resolution of about $1~\mathrm{kpc}$.
    
    The \ion{H}{i}~disc of NGC~55 appears to be distorted on all scales. There is a strong east--west asymmetry in the column density distribution along the major axis, suggesting that the disc is under the influence of ram-pressure forces. We also find evidence of streaming motions of the gas along the bar of NGC~55. The fitting of tilted rings to the velocity field reveals a strong warping of the outer gas disc which could be the result of tidal interaction with either NGC~300 or a smaller satellite galaxy. Finally, we find a large number of distinct clumps and spurs across the entire disc, indicating that internal or external processes, such as satellite accretion or gas outflows, have stirred up the gas disc.
    
    We also detect several isolated \ion{H}{i}~clouds within about $20~\mathrm{kpc}$ projected distance from NGC~55. Their dynamical properties and apparent concentration around NGC~55 suggest that most of the clouds are forming a circum-galactic population similar to the high-velocity clouds of the Milky Way and M31, although two of the clouds could be foreground objects and part of the Magellanic Stream. While it is difficult to determine the origin of these clouds, our data seem to favour either tidal stripping or gas outflows as the source of the gas.
  \end{abstract}
  
  
  \section{Introduction}
  
  The Sculptor group is one of the nearest galaxy groups beyond the Local Group, the distances of its members ranging from approximately $2$ to $5~\mathrm{Mpc}$ \citep{Jerjen1998}. At a distance of about $2~\mathrm{Mpc}$, the two nearest major group members, NGC~55 and NGC~300, are forming a distinct subgroup \citep{Karachentsev2003}. NGC~55 is a medium-sized, irregular, barred spiral galaxy of type SB(s)m \citep{deVaucouleurs1991} and closely resembles the Large Magellanic Cloud (LMC) in both morphology and luminosity. Unlike the LMC, NGC~55 appears almost edge-on, allowing more detailed studies of its stellar and gaseous halo while hampering studies of the morphology and kinematics of its disc. Optical studies indicate that the bar is seen almost end-on and slightly offset from the centre of the galaxy \citep{deVaucouleurs1961,Robinson1966}.
  
  Some of the basic properties and physical parameters of NGC~55, including the results of this work, are summarised in Table~\ref{tab_ngc55}. The distance to NGC~55 was recently measured by \citet{Pietrzynski2006} to be $1.91^{+0.13}_{-0.12}~\mathrm{Mpc}$, using the period-luminosity relationship of 143 newly discovered Cepheids. A slightly larger value of $2.30 \pm 0.35~\mathrm{Mpc}$ was found by \citet{vandeSteene2006}, using the luminosity function of 21~planetary nebulae discovered in NGC~55. Throughout this paper we adopt a distance of $1.9~\mathrm{Mpc}$ for NGC~55 based on the results by \citet{Pietrzynski2006}.
  
  \begin{table}
    \centering
    \caption{Properties of NGC~55. References: [1]~\citet{deVaucouleurs1991}; [2]~\citet{Jarrett2000}; [3]~\citet{Pietrzynski2006}; [4]~\citet{Dale2009}. Parameters without reference in the lower part of the table are the result of this study.}
    \label{tab_ngc55}
    \begin{tabular}{lrll}
      \hline
      Parameter & Value & Unit & Ref. \\
      \hline
      Type                     &                             SB(s)m &                               & [1] \\
      $\alpha$ (J2000)         &   $00^{\rm h} 14^{\rm m} 53\fs{}6$ &                               & [2] \\
      $\delta$ (J2000)         & ${-39}\degr{} 11\arcmin{} 47\farcs{}9$ &                           & [2] \\
      Distance                 &                      $1.9 \pm 0.1$ & $\mathrm{Mpc}$                & [3] \\
      Major axis ($D_{25}$)    &                     $32.4 \pm 0.7$ & $\mathrm{arcmin}$             & [1] \\
      Total IR luminosity      &                $1.6 \times 10^{9}$ & $\mathrm{L}_{\sun}$           & [4] \\
      Radial velocity          &                                    &                               & \\
      \quad barycentric        &                        $131 \pm 2$ & $\mathrm{km \, s}^{-1}$       & \\
      \quad LSR                &                        $125 \pm 2$ & $\mathrm{km \, s}^{-1}$       & \\
      Integrated flux$^{1}$    &      $(2.0 \pm 0.1) \times 10^{3}$ & $\mathrm{Jy \, km \, s}^{-1}$ & \\
      \ion{H}{i} mass$^{1}$    &      $(1.7 \pm 0.1) \times 10^{9}$ & $\mathrm{M}_{\sun}$           & \\
      Total mass$^{2}$         &     $(2.0 \pm 0.4) \times 10^{10}$ & $\mathrm{M}_{\sun}$           & \\
      Position angle$^{3,4}$   &                                    &                               & \\
      \quad inner disc$^5$     &          $109\fdg{}7 \pm 0\fdg{}5$ &                               & \\
      \quad outer disc         &   $107\degr{} \! \ldots 93\degr{}$ &                               & \\
      Inclination$^{4}$        &    $85\degr{} \! \ldots 65\degr{}$ &                               & \\
      Max. rot. velocity$^{4}$ &                     $90.6 \pm 2.5$ & $\mathrm{km \, s}^{-1}$       & \\
      \hline
      \multicolumn{4}{l}{\footnotesize $^{1}\,$lower limit due to optical depth and missing short spacings} \\
      \multicolumn{4}{l}{\footnotesize $^{2}\,$within a radius of $18~\mathrm{kpc}$ / $33~\mathrm{arcmin}$} \\
      \multicolumn{4}{l}{\footnotesize $^{3}\,$w.r.t.\ the J2000.0 equatorial coordinate system} \\
      \multicolumn{4}{l}{\footnotesize $^{4}\,$based on tilted-ring fit} \\
      \multicolumn{4}{l}{\footnotesize $^{5}\,$within a radius of $8.8~\mathrm{kpc}$ / $16~\mathrm{arcmin}$}
    \end{tabular}
  \end{table}
  
  A summary of previous \ion{H}{i}~observations of NGC~55 is given in Table~\ref{tab_hiobs}. Early \ion{H}{i}~observations of the galaxy were presented by \citet{Epstein1964}, using the Harvard 60-ft telescope, and by \citet{Robinson1964,Robinson1966} based on observations with the 64-m Parkes radio telescope. \citet{Robinson1964,Robinson1966} derived a total \ion{H}{i}~mass of $2.0 \times 10^{9}~\mathrm{M}_{\sun}$ (at an assumed distance of $1.74~\mathrm{Mpc}$) and a total dynamical mass of $2.5 \times 10^{10}~\mathrm{M}_{\sun}$ based on the measured rotation curve. The $14~\mathrm{arcmin}$~beam of the Parkes telescope, however, did not allow them to resolve details in the \ion{H}{i}~disc of the galaxy.
  
  First interferometric \ion{H}{i}~observations with the Owens Valley twin-element interferometer at a separation of about $30~\mathrm{m}$ were presented by \citet{Seielstad1965}, although the galaxy did not entirely fit into the primary beam of the 27-m telescopes. Their observations essentially confirmed the results obtained by \citet{Robinson1964}. They did not find any kinematic signature of streaming motions along the suspected optical bar of NGC~55 \citep{deVaucouleurs1961}.
  
  Interferometric \ion{H}{i}~observations (in addition to 6-cm radio continuum observations) at a much higher angular resolution of $50\arcsec{} \! \times 35\arcsec{}$ were carried out by \citet{Hummel1986} with the Very Large Array (VLA). Due to the limited field of view (single pointing) and sensitivity, their data only cover the inner, brighter regions of the \ion{H}{i}~disc of NGC~55 out to a column density level of just under $10^{20}~\mathrm{cm}^{-2}$. The total derived \ion{H}{i}~mass of $1.4 \times 10^{9}~\mathrm{M}_{\sun}$ at $1.9~\mathrm{Mpc}$ is somewhat smaller than the value obtained by \citet{Robinson1966}, reflecting the limited field of view of their observations in combination with the inevitable lack of short telescope spacings in interferometric observations.
  
  \begin{figure}
    \centering
    \includegraphics[width=0.85\linewidth]{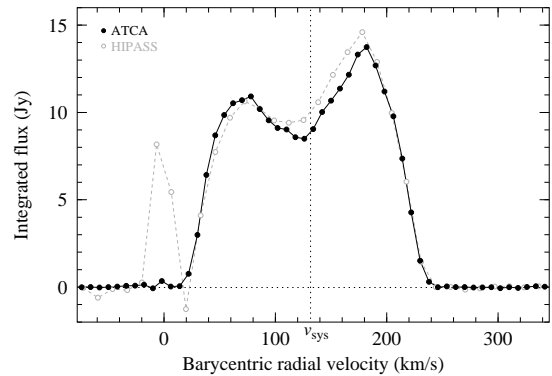}
    \caption{Integrated \ion{H}{i} spectrum of NGC~55 from our ATCA data (black, solid line) as compared to the integrated HIPASS spectrum (grey, dashed line) from the HIPASS Bright Galaxy Catalogue \citep{Koribalski2004}. Signals near $0~\mathrm{km \, s}^{-1}$ in both spectra are due to Galactic foreground emission.}
    \label{fig_intspec}
  \end{figure}
  
  NGC~55 was re-observed with the VLA by \citet{Puche1991} with a setup, angular resolution and sensitivity similar to \citet{Hummel1986}. Their total \ion{H}{i}~mass of $1.3 \times 10^{9}~\mathrm{M}_{\sun}$ at $1.9~\mathrm{Mpc}$ is comparable to the value found by \citet{Hummel1986}. By fitting a set of tilted rings to the velocity field, \citet{Puche1991} also derived the rotation curve of NGC~55 out to a radius of $22~\mathrm{arcmin}$ (approximately $12~\mathrm{kpc}$). They found a maximum rotation velocity of almost $90~\mathrm{km \, s}^{-1}$ (beyond a radius of $15~\mathrm{arcmin}$) and a dynamical mass of $1.8 \times 10^{10}~\mathrm{M}_{\sun}$ (at the outermost point of the rotation curve, assuming a distance of $1.6~\mathrm{Mpc}$).
  
  NGC~55 has been extensively studied at other wavelengths, including far-infrared \citep{Engelbracht2004}, optical and H$\upalpha$ \citep{Ferguson1996,Davidge2005,Castro2008,Tanaka2011}, and X-ray \citep{Schlegel1997,Stobbart2006}.
  
  In this paper we present the results of sensitive \ion{H}{i}~observations of a wide field around NGC~55 with the Australia Telescope Compact Array (ATCA) with the aim to study the extended gas disc of the galaxy and search for evidence of interaction and accretion. Our data supersede previous interferometric observations in the sense that we have imaged a much larger area of about $2\degr{} \! \times 2\degr{}$ around NGC~55 down to much lower \ion{H}{i}~column density levels of about $10^{19}~\mathrm{cm}^{-2}$, allowing us to trace faint gas in the vicinity of the galaxy and reveal the vast extent of the gas disc at this sensitivity and resolution for the first time.
  
  This paper is part of a series of papers describing deep \ion{H}{i}~studies of the major galaxies in the Sculptor group with the ATCA. The motivation, observations and data reduction strategy are described in more detail in the first paper of the series on NGC~300 (\citealt{Westmeier2011}; hereafter referred to as Paper~1).
  
  \begin{table*}
    \centering
    \caption{Summary of \ion{H}{i}~observations of NGC~55. $\vartheta$ denotes the angular resolution of the observation (FWHM of main beam or synthesised beam), $F_{\rm HI}$ is the integrated \ion{H}{i}~flux, and $M_{\rm HI}$ is the total \ion{H}{i}~mass scaled to a common distance of $1.9~\mathrm{Mpc}$.}
    \label{tab_hiobs}
    \begin{tabular}{lllrrr}
      \hline
      Reference              & Telescope              & Method              &         $\vartheta$ &   $F_{\rm HI}$ &  $M_{\rm HI}$ \\
                             &                        &                     &                     & ($\mathrm{Jy \, km \, s}^{-1}$) & ($10^{9}~M_{\rm \sun}$) \\
      \hline
      \citet{Epstein1964}    & Harvard 60~ft          & Drift-scan          &       $53\arcmin{}$ & $2450 \pm 250$ & $2.1 \pm 0.2$ \\
      \citet{Seielstad1965}  & Owens Valley interfer. & Single pointing     &   $24\arcmin{}^{1}$ &       --$^{2}$ &      --$^{2}$ \\
      \citet{Robinson1966}   & Parkes 64~m            & Map of 60~point.    &       $13\farcm{}5$ & $2800$         & $2.4$         \\
      \citet{Hummel1986}     & VLA                    & Single pointing     &  $50\arcsec{} \! \times 35\arcsec{}$ & $1600 \pm 100$ & $1.4 \pm 0.1$ \\
      \citet{Puche1991}      & VLA                    & Mosaic of 2~point.  &              $45\arcsec{}$ & $1520$         & $1.3$         \\
      \citet{Koribalski2004} & Parkes 64~m            & Map (HIPASS)        &       $15\farcm{}5$ & $1990 \pm 150$ & $1.7 \pm 0.1$ \\
      This work              & ATCA                   & Mosaic of 32~point. & $158\arcsec{} \! \times 84\arcsec{}$ & $1980 \pm 100$ & $1.7 \pm 0.1$ \\
      \hline
      \multicolumn{6}{l}{$^{1}\,$\footnotesize Zenith fringe spacing \citep{Clark1965}.} \\
      \multicolumn{6}{l}{$^{2}\,$\footnotesize Not specified, as the galaxy did not fit within the $30\arcmin{}$ primary beam of the telescope.}
    \end{tabular}
  \end{table*}
  
  \section{Observations}
  
  The observing strategy is identical to the one for NGC~300 and described in great detail in Paper~1. Here, we briefly summarise the main observing parameters.
  
  NGC~55 was observed with the Australia Telescope Compact Array between 2007 November and 2009 February in a mosaic of 32~pointings with a total area of approximately $2\degr{} \! \times 2\degr{}$. At a distance of $1.9~\mathrm{Mpc}$ this corresponds to a projected field size of about $65~\mathrm{kpc}$. The total integration time of $96~\mathrm{h}$ was equally divided among the EW352 and EW367 array configurations to improve spatial frequency coverage. Because of the large spatial separation between antenna~6 and all other antennas in these configurations, we only used data from antennas~1 to~5 in our analysis. The resulting minimum and maximum baselines were $31$ and $367~\mathrm{m}$, respectively.
  
  To improve the sensitivity even further, we added some more data of NGC~55 taken with the ATCA as part of the Local Volume \ion{H}{i} Survey (LVHIS; \citealt{Koribalski2010}). The data were observed in July and October~2005 in the H75 and EW214 array configurations, respectively, thus improving our surface brightness sensitivity by adding a range of short baselines to our EW352 and EW367 data. The LVHIS mosaics consist of 19~individual pointings covering an area of about $2\fdg{}5 \times 1\fdg{}5$ across the central part of our mosaic. Again, only data from antennas~1 to~5 were used.
  
  \begin{figure*}
    \centering
    \includegraphics[width=\linewidth]{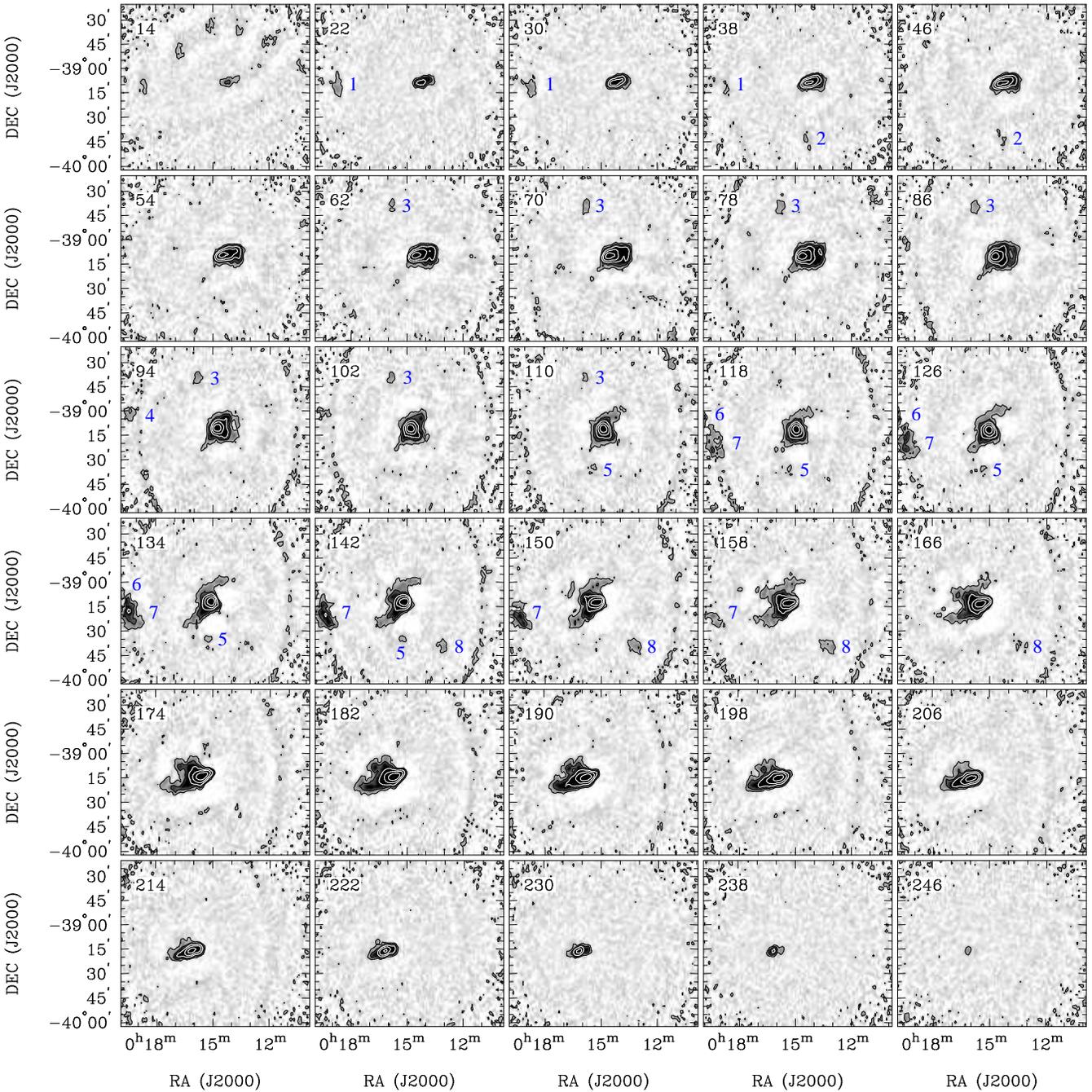}
    \caption{Channel maps of the ATCA \ion{H}{i} data cube of NGC~55 in the barycentric velocity range of $v_{\rm bar} = 14$ to $246~\mathrm{km \, s}^{-1}$ (corresponding to LSR velocities of $8$ to $240~\mathrm{km \, s}^{-1}$). The grey-scale images range from $-5$ to $+50~\mathrm{mJy}$ per beam. The contour levels are $9$, $27$, $81$, $243$, and $729~\mathrm{mJy}$ per beam (corresponding to $0.4$, $1.2$, $3.7$, $11.1$, and $33.3~\mathrm{K}$), and the beam size is $158 \times 84~\mathrm{arcsec}$. The increased noise along the edges of the maps is the result of primary beam attenuation. The isolated gas clouds identified in the data cube have been labelled with blue numbers.}
    \label{fig_chanmaps}
  \end{figure*}
  
  \section{Data reduction}
  
  Our data reduction strategy is explained in great detail in Paper~1, and exactly the same procedure was applied to NGC~55, again using PKS~1934$-$638 and PKS~0008$-$421 as the primary and secondary calibrator, respectively.
  
  The final spectral baseline rms of the NGC~55 data cube is $3.0~\mathrm{mJy}$ per beam at $8~\mathrm{km \, s}^{-1}$ velocity resolution and channel width (after spectral smoothing). Using robust weighting of the visibility data with a robustness parameter of~$0$, the synthesised beam has a full width at half maximum (FWHM) of $158 \times 84~\mathrm{arcsec}$, corresponding to a physical scale of $1.5 \times 0.8~\mathrm{kpc}$ at the distance of NGC~55 (at $d = 1.9~\mathrm{Mpc}$, $1~\mathrm{arcmin}$ corresponds to $0.55~\mathrm{kpc}$). The resulting $5 \sigma$ \ion{H}{i}~column density sensitivity per spectral channel for emission filling the synthesised beam is $1 \times 10^{19}~\mathrm{cm}^{-2}$. This corresponds to about $1.9 \times 10^{19}~\mathrm{cm}^{-2}$ across a typical line width of $30~\mathrm{km \, s}^{-1}$ FWHM.
  
  \section{Results}
  
  \subsection{NGC~55}
  \label{sect_ngc55}
  
  Fig.~\ref{fig_intspec} shows the integrated \ion{H}{i} spectrum of NGC~55. The spectrum is clearly asymmetric with stronger emission from the south-eastern, receding side of the galaxy. This is consistent with previous observations, e.g.\ by \citet{Puche1991}.\footnote{Please note that the velocity axis labels and designation of the south-eastern and north-western side of NGC~55 in figure~5 of \citet{Puche1991} are wrong. The velocity of $0~\mathrm{km \, s}^{-1}$ should be on the right-hand side of their spectrum (also see Appendix~\ref{app_puche}).} From the first moment of the spectrum we derive a systemic velocity of NGC~55 of $v_{\rm sys} = 131 \pm 2~\mathrm{km \, s}^{-1}$ in the barycentric reference frame, equivalent to a local standard-of-rest (LSR) velocity of $125 \pm 2~\mathrm{km \, s}^{-1}$. This is consistent with the barycentric velocity of $130~\mathrm{km \, s}^{-1}$ measured by \citet{Robinson1966} and the value of $129 \pm 2~\mathrm{km \, s}^{-1}$ found by \citet{Koribalski2004} based on the midpoint of the 50~per cent level of the peak flux density in the \ion{H}{i} Parkes All-Sky Survey (HIPASS; \citealt{Barnes2001}). Our velocity measurement is inconsistent with the barycentric velocity of $v_{\rm sys} = 118.3 \pm 4.0~\mathrm{km \, s}^{-1}$ derived by \citet{Puche1991}. This inconsistency is presumably the result of an error in the data analysis by \citet{Puche1991} and investigated in more detail in Appendix~\ref{app_puche}.
  
  \begin{figure*}
    \centering
    \includegraphics[width=0.85\linewidth]{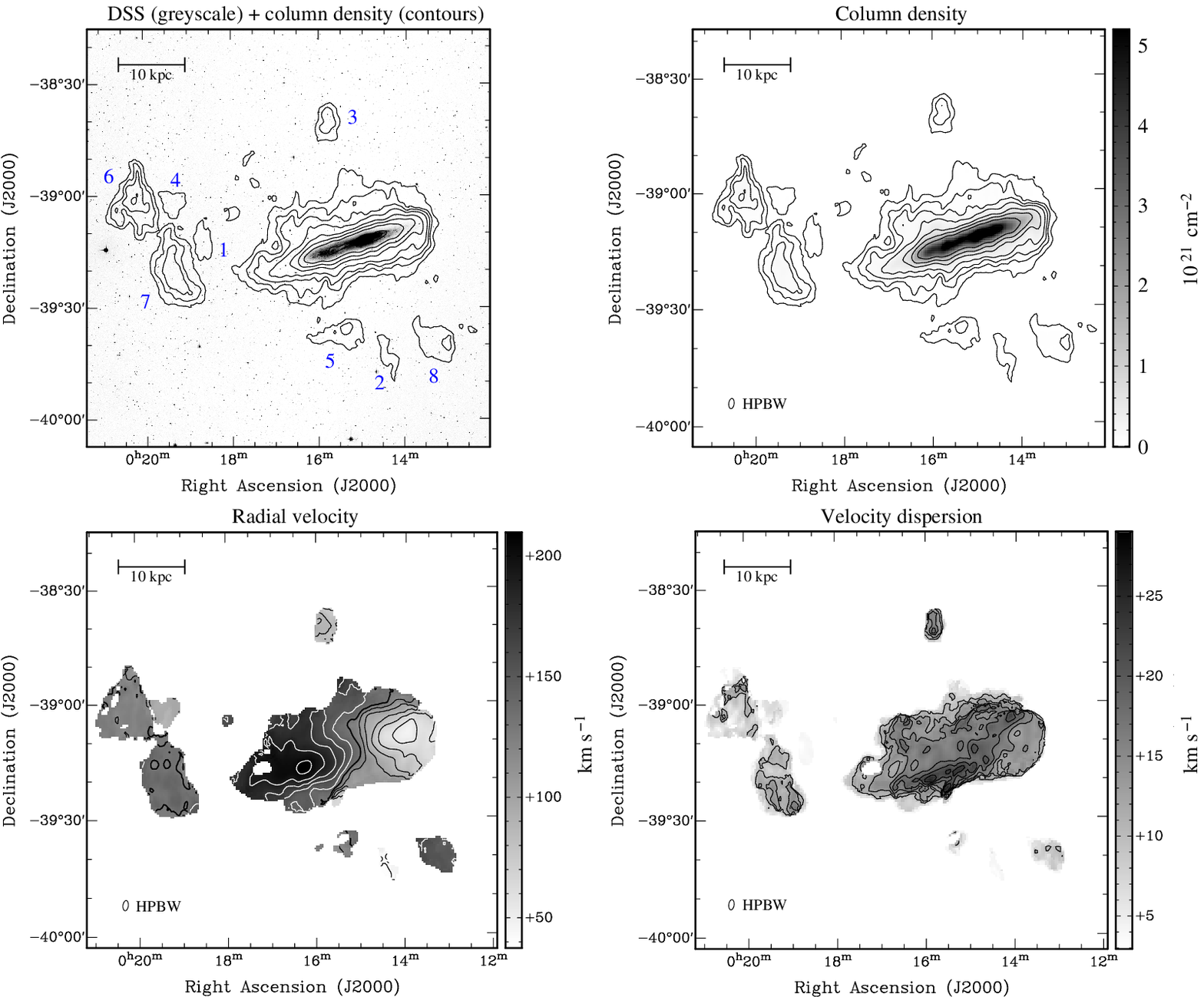}
    \caption{Top-left panel: Optical image of NGC~55 from the Digitized Sky Survey with \ion{H}{i} column density contours overlaid. The isolated gas clouds are labelled with blue numbers corresponding to those in Table~\ref{tab_hvcs}. Top-right panel: \ion{H}{i} column density map of NGC~55 derived from the zeroth moment under the assumption that the gas is optically thin. The black \ion{H}{i} column density contours in this and the previous map correspond to $0.1$, $0.5$, $1$, $2$, $5$, $10$, $20$, and $50 \times 10^{20}~\mathrm{cm}^{-2}$. Bottom-left panel: Barycentric radial velocity map derived from the first moment. The contours are drawn at intervals of $15~\mathrm{km \, s}^{-1}$ centred on the systemic velocity of NGC~55 of $v_{\rm sys} = 131~\mathrm{km \, s}^{-1}$ (bold contour line). Bottom-right panel: Velocity dispersion map derived from the second moment. The contours correspond to $10$, $12.5$, $15$, $17.5$, $20$, and $22.5~\mathrm{km \, s}^{-1}$. The black ellipse in each map indicates the beam size of $158 \times 84~\mathrm{arcsec}$.}
    \label{fig_ngc55}
  \end{figure*}
  
  We measure an integrated \ion{H}{i}~flux of $1980 \pm 100~\mathrm{Jy \, km \, s}^{-1}$ for NGC~55. At a distance of $1.9~\mathrm{Mpc}$ this corresponds to a total \ion{H}{i} mass of $(1.7 \pm 0.1) \times 10^{9}~\mathrm{M}_{\sun}$. This value has not been corrected for optical depth effects and missing diffuse flux and must therefore be considered a lower limit. In comparison, \citet{Puche1991} derived a total \ion{H}{i} mass of $1.3 \times 10^{9}~\mathrm{M}_{\sun}$ from their VLA observations (scaled to a distance of $1.9~\mathrm{kpc}$), and the single-dish \ion{H}{i} mass from the HIPASS Bright Galaxy Catalogue is $(1.7 \pm 0.1) \times 10^{9}~\mathrm{M}_{\sun}$ \citep{Koribalski2004}, suggesting that we picked up a large fraction of the total flux of NGC~55 in our ATCA observations. A larger value of $2.4 \times 10^{9}~\mathrm{M}_{\sun}$ (scaled to a distance of $1.9~\mathrm{Mpc}$) was found by \citet{Robinson1966} from observations with the Parkes radio telescope, indicating that the HIPASS flux may be slightly too low, possibly as the result of bandpass artefacts near Galactic velocities in HIPASS. The integrated HIPASS spectrum of NGC~55 is plotted in Fig.~\ref{fig_intspec} for comparison.
  
\begin{figure}
    \centering
    \includegraphics[width=0.85\linewidth]{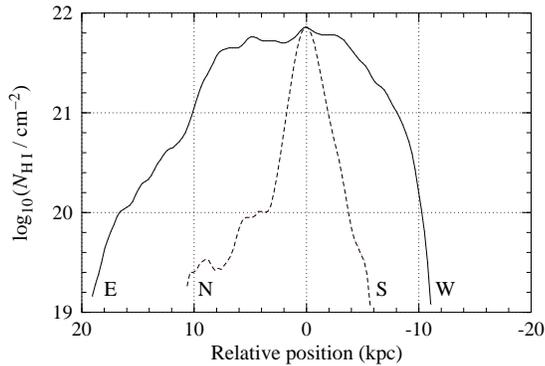}
    \caption{\ion{H}{i} column density profile of NGC~55 along the major axis (solid curve) and minor axis (dashed curve) through the point of peak column density. The capital letters indicate the orientation on the sky.}
    \label{fig_profiles}
  \end{figure}

  The individual channel maps from the data cube are shown in Fig.~\ref{fig_chanmaps}, while a map of the total \ion{H}{i} column density of NGC~55, calculated on the basis of the zeroth moment of the cube, is shown in the two upper panels of Fig.~\ref{fig_ngc55}. Comparison with an optical image from the Digitized Sky Survey (upper-left panel of Fig.~\ref{fig_ngc55}) demonstrates that the higher column density contours of $N_{\rm H\,I} \ga 10^{21}~\mathrm{cm}^{-2}$ near the disc plane of NGC~55 closely follow the stellar light distribution of the galaxy. At lower column density levels, however, NGC~55 is very extended, showing a complex distribution of `extra-planar' gas as well as several isolated \ion{H}{i} clouds. The overall column density distribution is clearly asymmetric with a sudden and sharp drop in column density on the western side of the disc (right-hand side in Fig.~\ref{fig_ngc55}), but a rather extended `tail' on the eastern side. Similarly, the gas disc appears to be more extended on the northern side of the disc than in the south. These effects are illustrated in Fig.~\ref{fig_profiles} which shows the column density profiles along the major and minor axis of NGC~55. Assuming an exponential column density profile of the form $\log_{10}(N_{\rm H\,I} / \mathrm{cm}^{-2}) = a (x / \mathrm{kpc}) + b$, we derive a slope of $a = 0.18$ for the eastern side of the disc, whereas on the western side the column density appears to decrease faster than exponentially.

\begin{figure}
    \centering
    \includegraphics[width=\linewidth]{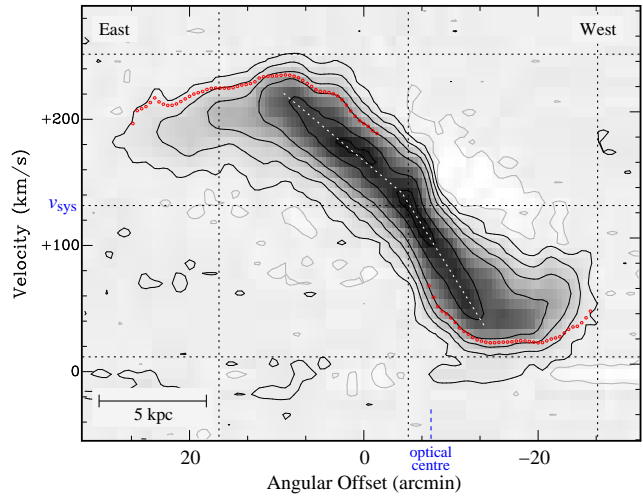}
    \caption{Position--velocity map along the major axis of NGC~55 (centre position: $\alpha = 0^{\rm h}15^{\rm m}33^{\rm s}$, $\delta = -39\degr{}11'52''$; position angle: $109\degr{}$). The contour levels are $-0.005$ (grey), $0.005$, $0.03$, $0.1$, $0.3$, $1$, and $1.5~\mathrm{Jy}$ per beam. Positive and negative signals near $v_{\rm bar} \approx 0~\mathrm{km \, s}^{-1}$ are due to Galactic foreground, and areas of `negative flux' near NGC~55 are the result of missing short spacings. The dotted, horizontal and vertical grid lines are equidistant and have been inserted to guide the eye and illustrate the substantial asymmetries within the gas disc of NGC~55. The small, red symbols (circles) mark the terminal velocities derived by the envelope-tracing method.}
    \label{fig_posivelo}
  \end{figure}

  This conspicuous asymmetry, already noted by \citet{Robinson1964}, is also obvious from the position--velocity map along the major axis of the gas disc of NGC~55 which is shown in Fig.~\ref{fig_posivelo}. At almost all flux contour levels, and in particular at the lowest detected flux levels, the eastern half of the galaxy is substantially more extended than the western half. At the same time there is a noticeable kink in the position--velocity diagram near the centre of NGC~55, the right-hand side being considerably steeper than the left-hand side, as indicated by the white, dotted line which was included for the purpose of guiding the eye.
  
  The radial velocity map of NGC~55 is shown in the bottom-left panel of Fig.~\ref{fig_ngc55} and reflects some of the asymmetries discussed before. Along the major axis the velocity gradient is larger across the western half of the galaxy than across the eastern half, indicated by smaller separations between the velocity contours. This asymmetry is related to the aforementioned kink in the position--velocity diagram. Along the minor axis the velocity contours are not straight, but s-shaped. This s-shape is also apparent from the individual channel maps in Fig.~\ref{fig_chanmaps} for velocities near the systemic velocity, suggesting that the extended emission perpendicular to the major axis is not due to halo gas located above the disc, but rather due to strong warping of the outer \ion{H}{i}~disc of NGC~55, similar to NGC~300 (Paper~1).
  
  The bottom-right panel of Fig.~\ref{fig_ngc55} shows the radial velocity dispersion across the disc of NGC~55 as derived from the second moment of the \ion{H}{i}~spectra. The typical dispersion across most of the inner disc is about $15~\mathrm{km \, s}^{-1}$, corresponding to a FWHM of about $35~\mathrm{km \, s}^{-1}$. In the region of the bar, however, we find a slightly higher dispersion of about $17$ to $20~\mathrm{km \, s}^{-1}$, presumably reflecting the larger velocity gradient in that region due to the kink in the position--velocity diagram. Even higher dispersions in excess of $20~\mathrm{km \, s}^{-1}$ are seen along the southern edge of NGC~55, possibly resulting from different layers of gas along the line of sight due to the warping of the outer \ion{H}{i}~disc. The highest dispersions of up to about $30~\mathrm{km \, s}^{-1}$ are found in the region of the southern spur, a conspicuous spur of gas that is particularly prominent in the individual channel maps in Fig.~\ref{fig_chanmaps} in the velocity range of about $80$ to $100~\mathrm{km \, s}^{-1}$.
  
  In summary, the \ion{H}{i}~disc of NGC~55 appears distorted, suggesting that the galaxy has recently been subjected to internal or external processes (such as tidal interaction, accretion, ram pressure, or star formation) that have led to the observed asymmetries and distortions. In Section~\ref{sect_discussion} we will assess in more detail the different potential processes that could have shaped the \ion{H}{i}~disc.
  
  \begin{table*}
    \centering
    \caption{Parameters of the isolated \ion{H}{i} clouds (1--8) and the north-eastern clump (NEC) found in the NGC~55 field. Columns~2 and~3 denote the column-density-weighted mean position (right ascension, $\alpha$, and declination, $\delta$). Columns~4 to~6 denote the barycentric radial velocity, $v_{\rm bar}$, velocity dispersion, $\sigma_{v}$, and total \ion{H}{i} mass, $M_{\rm HI}$, based on fitting a Gaussian to the integrated spectrum of each cloud (assuming $d = 1.9~\mathrm{Mpc}$). Columns~7 to~9 denote the peak brightness temperature, $T_{\rm B}$, peak \ion{H}{i}~column density, $N_{\rm H\,I}$, and a rough estimate of density, $n_{\rm H\,I} = N_{\rm H\,I} / D$, under the assumption of a constant diameter of $D = 2~\mathrm{kpc}$. Column~10 denotes the cloud's diameter along the major and minor axis based on the fitting of a two-dimensional, elliptical Gaussian function to the \ion{H}{i}~column density map.}
    \label{tab_hvcs}
    \begin{tabular}{rrrrrrrrrr}
      \hline
        Cloud &   $\alpha$ &       $\delta$ &   $v_{\rm bar}$ &   $\sigma_{v}$ &   $M_{\rm HI}$ & $T_{\rm B}$ & $N_{\rm H\,I}$ & $\log n_{\rm H\,I}$ & $D_{\rm maj} \times D_{\rm min}$ \\
            &    (J2000) &        (J2000) & ($\mathrm{km \, s}^{-1}$) & ($\mathrm{km \, s}^{-1}$) & ($10^{6}~\mathrm{M}_{\sun}$) & (K) & ($10^{19}~\mathrm{cm^{-2}}$) & ($\mathrm{cm^{-3}}$) & ($\mathrm{kpc}$) \\
      \hline
        1 & 00:18:41.2 & $-39$:12:07 &  $21.3 \pm 1.8$ & $16.0 \pm 1.8$ & $ 7.2 \pm 1.6$ & $1.30 \pm 0.14$ &  $4.6 \pm 1.0$ & $-2.1$ & $4.9 \times 2.2$ \\
        2 & 00:14:24.2 & $-39$:42:58 &  $41.0 \pm 1.4$ & $ 8.1 \pm 1.4$ & $ 2.2 \pm 0.7$ & $0.93 \pm 0.14$ &  $2.7 \pm 0.9$ & $-2.4$ & $7.0 \times 1.9$ \\
        3 & 00:15:47.5 & $-38$:40:20 &  $86.8 \pm 2.1$ & $22.8 \pm 2.1$ & $ 7.7 \pm 1.3$ & $1.19 \pm 0.14$ & $11.1 \pm 1.6$ & $-1.7$ & $3.6 \times 2.1$ \\
        4 & 00:19:21.8 & $-39$:02:04 &  $94.4 \pm 1.8$ & $10.5 \pm 1.8$ & $ 3.5 \pm 1.1$ & $1.04 \pm 0.14$ &  $4.1 \pm 1.2$ & $-2.2$ & $3.5 \times 2.5$ \\
        5 & 00:15:22.8 & $-39$:37:01 & $122.6 \pm 2.0$ & $22.9 \pm 1.9$ & $ 6.8 \pm 1.1$ & $0.75 \pm 0.14$ &  $6.0 \pm 1.2$ & $-2.0$ & $5.0 \times 1.5$ \\
        6 & 00:20:13.2 & $-39$:01:55 & $123.7 \pm 0.3$ & $ 8.5 \pm 0.3$ & $33.0 \pm 2.5$ & $7.16 \pm 0.14$ & $21.8 \pm 1.8$ & $-1.5$ & $5.4 \times 3.8$ \\
        7 & 00:19:19.1 & $-39$:20:42 & $136.7 \pm 0.4$ & $13.3 \pm 0.4$ & $36.1 \pm 2.0$ & $5.10 \pm 0.14$ & $15.1 \pm 1.2$ & $-1.6$ & $7.5 \times 3.4$ \\
        8 & 00:13:11.8 & $-39$:39:52 & $150.2 \pm 2.2$ & $17.6 \pm 2.2$ & $ 8.4 \pm 2.0$ & $1.36 \pm 0.14$ &  $6.8 \pm 1.0$ & $-2.0$ & $5.5 \times 3.1$ \\
      NEC & 00:16:41.0 & $-39$:09:37 & $183.8 \pm 0.4$ & $17.9 \pm 0.4$ & $12.6 \pm 0.5$ & $2.75 \pm 0.14$ & $20.4 \pm 0.7$ & $-1.5$ & $3.3 \times 2.2$ \\
      \hline
    \end{tabular}
  \end{table*}

  \subsection{Structures in the outer disc}
  \label{sect_outerdisc}
  
\begin{figure}
    \centering
    \includegraphics[width=\linewidth]{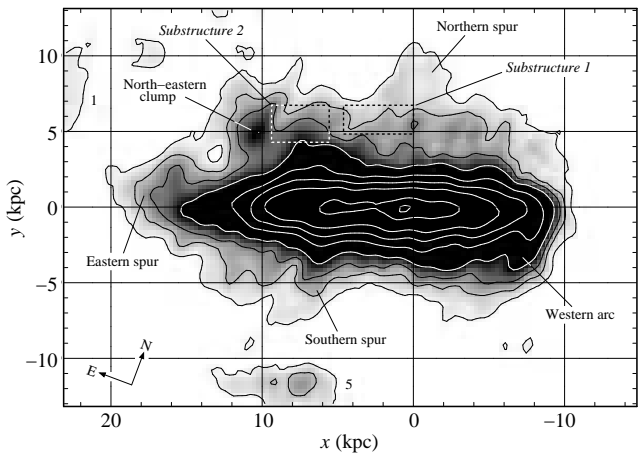}
    \caption{\ion{H}{i}~column density map of NGC~55 in coordinates of physical length relative to the optical centre of the galaxy. The map was rotated by $20\degr{}$ for alignment with the major axis of NGC~55. The linear grey-scale image ranges from $0$ to $2 \times 10^{20}~\mathrm{cm}^{-1}$. The contour levels are $0.1$, $0.5$, $1$, $2$, $5$, $10$, $20$, $50$, and $70 \times 10^{20}~\mathrm{cm}^{-1}$. The rectangles mark the regions of the stellar substructures in figure~4 of \citet{Tanaka2011}.}
    \label{fig_disc}
  \end{figure}

  Fig.~\ref{fig_disc} shows an \ion{H}{i}~column density map of the disc of NGC~55 with the most prominent features of the outer gas disc labelled. Some of these features were already identified by \citet{Hummel1986}, namely the conspicuous southern spur and western arc as well as the extended regions of gas in the north-east and south-west that are due to the warping of the outer \ion{H}{i}~disc (also see Section~\ref{sect_rotationcurve}). The high surface brightness sensitivity of our observations reveals several additional features for the first time, in particular a conspicuous clump in the north-east, an extended disc component and spur in the north and north-west, and another extended spur at the eastern end of the disc.
  
\begin{figure*}
    \centering
    \includegraphics[width=0.85\linewidth]{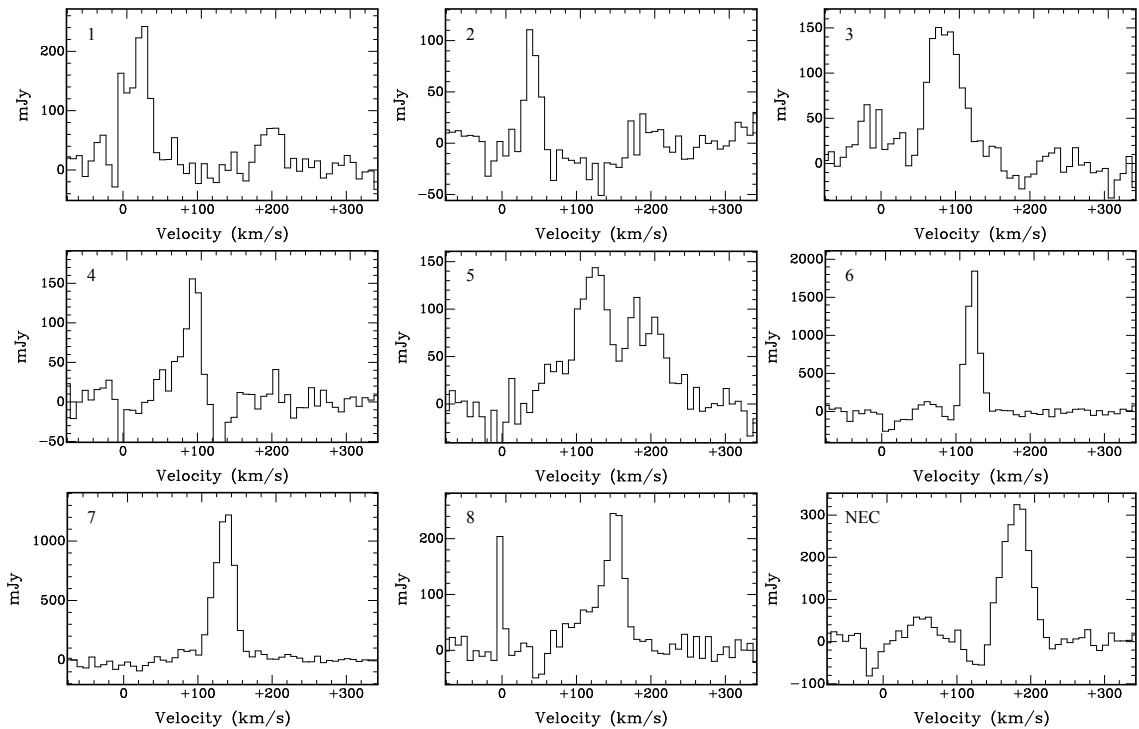}
    \caption{Integrated spectra of the isolated gas clouds (1--8) and the north-eastern clump (NEC) found in the field around NGC~55. The numbers correspond to those in the top-left panel of Fig.~\ref{fig_ngc55} and Table~\ref{tab_hvcs}. The positive and negative signals near $v_{\rm bar} \approx 0~\mathrm{km \, s}^{-1}$ are due to Galactic foreground emission.}
    \label{fig_hvc-spectra}
  \end{figure*}

  The north-eastern clump is remarkable in that it has a fairly high peak column density of $2 \times 10^{20}~\mathrm{cm}^{-2}$ and appears largely isolated in phase space from the disc of NGC~55. Its parameters are listed in Table~\ref{tab_hvcs}, and its integrated spectrum is presented in Fig.~\ref{fig_hvc-spectra}. The north-eastern clump is characterised by a fairly large velocity dispersion of $17.9 \pm 0.4~\mathrm{km \, s}^{-1}$, equivalent to a FWHM of about $42~\mathrm{km \, s}^{-1}$. The \ion{H}{i}~mass of the clump is approximately $1.3 \times 10^{7}~\mathrm{M}_{\sun}$, making it about a factor of two more massive than most of the other, isolated \ion{H}{i}~clouds near NGC~55. The clump may be more massive even than high-velocity cloud complex~C near the Milky Way which has an \ion{H}{i}~mass in the range of $0.3$ to $1.4 \times 10^{7}~\mathrm{M}_{\sun}$ \citep{Wakker2007,Thom2008}. Despite its high mass and column density, the clump does not appear to contain any stars. Deep optical imaging in the $V$ and $I$~bands by \citet{Tanaka2011} does not show any obvious excess of RGB stars towards the centre of the clump, although a faint stellar structure, referred to as substructure~2 by \citet{Tanaka2011}, is found nearby (Fig.~\ref{fig_disc}).
  
  Another remarkable feature is the southern spur which can be traced all the way from the inner disc of NGC~55 out to the faintest column density contour in Fig.~\ref{fig_disc} and is particularly prominent in the individual channel maps in the velocity range of about $80$ to $100~\mathrm{km \, s}^{-1}$ (Fig.~\ref{fig_chanmaps}). The inner parts of the spur were already detected in previous \ion{H}{i}~observations \citep{Hummel1986,Puche1991}, but our sensitive observations reveal that the feature can be traced out to a projected distance of about $7~\mathrm{kpc}$ from the major axis of NGC~55. In addition, we discovered an isolated (at the $10^{19}~\mathrm{cm}^{-2}$ column density level) \ion{H}{i}~cloud, number~5 in Table~\ref{tab_hvcs}, near the end of the spur. The radial velocities observed across cloud number~5 are similar to those near the end of the southern spur, suggesting that the cloud might be associated with the spur, although we failed to find convincing evidence for a connecting gas bridge between the two within the sensitivity of our data. While \citet{Puche1991} do not discuss the origin of the southern spur, \citet{Hummel1986} speculate that this and other peculiar features of the \ion{H}{i}~disc could be the result of interaction with NGC~300 and an `anonymous dwarf dI', by which they possibly refer to ESO~294$-$010 \citep{Karachentsev2003}.
  
  Similar extended spurs can also be found in the north and east, although they are not quite as narrow and confined as the southern spur. The northern spur is rather broad and faint with column densities of $5 \times 10^{19}~\mathrm{cm}^{-2}$ or less. It can be traced out to a projected distance of about $10~\mathrm{kpc}$ from the major axis of NGC~55. As with its southern counterpart, the northern spur is pointing towards an isolated gas cloud (number~3 in Table~\ref{tab_hvcs}). In this case, however, the cloud has a larger projected separation from the tip of the spur, and the radial velocities across the cloud ($v_{\rm bar} \approx 85~\mathrm{km \, s}^{-1}$) are rather different from the velocities found near the tip of the spur (approximately $175$ to $185~\mathrm{km \, s}^{-1}$), suggesting that the two structures are not related.
  
  The eastern spur is unique in that it appears to be a tail-like extension of the disc of NGC~55. As mentioned earlier, the eastern end of the disc is much more extended and less sharply defined than the western end. A detailed inspection of the individual channel maps in Fig.~\ref{fig_chanmaps} reveals that the eastern spur is not a continuous structure. In fact, the entire eastern and north-eastern end of the disc seems to be highly chaotic and fragmented, dispersing into an aggregation of indistinct clouds and filaments that are too faint to be individually classified (also see Fig.~\ref{fig_ngc55}).
  
  The multitude of irregular features detected in the outer disc of NGC~55 consolidates our impression of recent, strong distortions of the galaxy's gas disc by internal or external effects yet to be determined. A more detailed assessment of the potential origins of the observed irregularities will be attempted in Section~\ref{sect_discussion}.
  
  \subsection{Isolated gas clouds}
  \label{sect_hvcs}
  
  We found a population of isolated \ion{H}{i}~clouds in the field around NGC~55 with typical projected separations from the disc of about $10$ to $20~\mathrm{kpc}$. The top panels of Fig.~\ref{fig_ngc55} show a column density map of the clouds, and integrated spectra of all clouds are presented in Fig.~\ref{fig_hvc-spectra}. The basic physical parameters of the clouds are summarised in Table~\ref{tab_hvcs}. The first question that arises is whether these detections are real or whether they have been caused by either noise or deconvolution artefacts in the data. From the integrated spectra in Fig.~\ref{fig_hvc-spectra} it is evident that most of the clouds have been detected at a significant signal-to-noise ratio and are unlikely to have been caused by noise. Furthermore, their morphology and orientation within the three-dimensional data cube are inconsistent with their being the result of residual sidelobes of NGC~55 or artefacts of the deconvolution process.
  
  A few of the clouds found in our ATCA data were detected in previous surveys. This is the case for clouds number~6, 7, and~8 which were detected in a wide-field \ion{H}{i}~survey of the Sculptor group by \citet{Haynes1979}. Earlier observations carried out by \citet{Mathewson1975} with the 64-m Parkes telescope had uncovered seven \ion{H}{i}~detections in the vicinity of NGC~55. However, there appears to be little agreement between their map and ours, calling the genuineness of their detections into question.
  
\begin{figure*}
    \centering
    \includegraphics[width=0.85\linewidth]{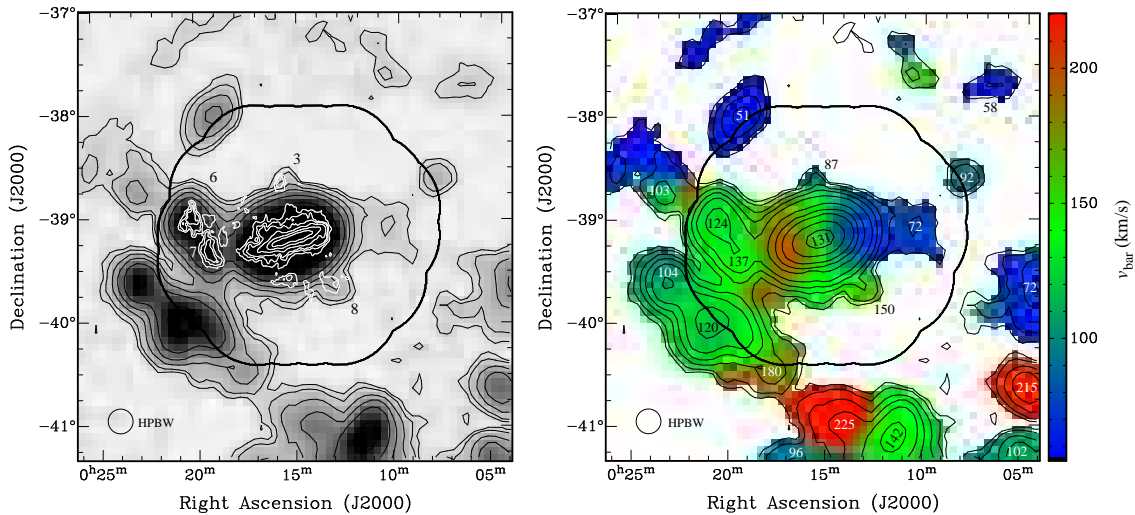}
    \caption{HIPASS \ion{H}{i}~column density map (left) and radial velocity map (right) of the region around NGC~55 in the barycentric velocity range of $52$ to $250~\mathrm{km \, s}^{-1}$. The contour levels are $0.1$, $0.2$, $0.5$, $1$, $2$, $5$, $10$, $20$, $50$, and $100 \times 10^{19}~\mathrm{cm}^{-2}$. The black circle in the bottom-left corner indicates the half-power beam width of HIPASS. The \ion{H}{i} column density map based on our ATCA data is plotted in the left-hand panel as the white contours at levels of $0.1$, $0.5$, $1$, $3$, $10$, and $30 \times 10^{20}~\mathrm{cm}^{-2}$, demonstrating that several of the gas clouds near NGC~55, in particular number~3, 6, 7, and 8, are clearly detected in HIPASS. The bold, black line outlines the edge of the ATCA mosaic. The numbers in the right-hand panel specify barycentric radial velocities in $\mathrm{km \, s}^{-1}$ of individual features.}
    \label{fig_hipass}
  \end{figure*}

  The most convincing piece of evidence for the existence of the clouds is provided by the \ion{H}{i}~Parkes All-Sky Survey (\citealt{Barnes2001}). In Fig.~\ref{fig_hipass} we present a moment~0 map of the region around NGC~55 from HIPASS with \ion{H}{i} column density contours from our ATCA data overlaid. Those clouds that are sufficiently separated from NGC~55 to not blend in with the bright emission from the galaxy itself are clearly detected in HIPASS at the correct position and velocity. This is the case in particular for clouds~3 and~8, and for the two bright and extended clouds~6 and~7 in the left-hand part of the map. Those detections must therefore be genuine objects, and it seems likely that the remaining clouds, which are too close to NGC~55 to be individually detected in HIPASS, are genuine as well.
  
  The next question we need to address is whether the gas clouds are indeed forming a circum-galactic population around NGC~55 or whether they constitute foreground objects in the vicinity of the Milky Way. There is good reason to consider the possibility of a Galactic nature of the clouds, given that the radial velocity range of NGC~55 partly overlaps with that of the Milky Way in this direction. To complicate matters further, the Magellanic Stream is running across the same part of the sky, and its velocities again overlap with both the Milky Way and NGC~55. In fact, the HIPASS column density and velocity maps in Fig.~\ref{fig_hipass} reveal additional emission beyond the boundaries of our ATCA mosaic which is most likely due to Galactic or Magellanic foreground. Several of the low-velocity features at $v_{\rm bar} \la 70~\mathrm{km \, s}^{-1}$ are most likely associated with the Milky Way, while gas clouds presumably belonging to the Magellanic Stream can be found across a wide range of velocities from about $100$ to $225~\mathrm{km \, s}^{-1}$. Therefore, we are confronted with a complete spatial and kinematic overlap between foreground emission and potential structures associated with NGC~55 and the Sculptor group.
  
  A first indication of the true nature of the isolated gas clouds found in our ATCA data comes from their distribution in the sky. Although we have mapped a large area on the sky, Fig.~\ref{fig_ngc55} and~\ref{fig_hipass} reveal that the entire population of clouds is highly concentrated around NGC~55. Across large parts of the ATCA mosaic at larger angular separations from NGC~55 there are no detections at all, although the noise levels are similar to those near the centre of the mosaic and would have allowed us to detect gas clouds out to much larger angular distances from NGC~55.

  Two objects, clouds~6 and~7, differ from the rest of the population by having substantially larger \ion{H}{i}~masses than any other cloud (see Table~\ref{tab_hvcs}). At the same time, their spectral lines are among the narrowest encountered among the isolated gas clouds, while we would have expected the line width to increase with mass.
  
  This result leads us to conclude that clouds~1--5 and~8 likely form a circum-galactic population in the vicinity of NGC~55, whereas clouds~6 and~7 are most likely foreground objects, presumably associated with the Magellanic Stream which has similar radial velocities in this part of the sky. This is also consistent with the HIPASS radial velocity map in the right-hand panel of Fig.~\ref{fig_hipass}, which illustrates that there is a more widely distributed population of extended clouds with velocities between about $100$ and $150~\mathrm{km \, s}^{-1}$ which are presumably part of the Magellanic Stream that runs across the region. In fact, clouds~6 and~7 do not appear to be isolated objects, but part of a complex of clouds with similar velocities and properties.
  
  \section{Rotation curve and mass}
  \label{sect_rotationcurve}
  
  Determining the rotation curve of NGC~55 is important for the accurate measurement of its dynamical mass and for studies of its mass composition and dark-matter content (flat versus declining rotation curve). Attempts to measure the rotation curve of NGC~55 have been hampered by the galaxy's high inclination of approximately $80\degr{}$ \citep{Hummel1986,KiszkurnoKoziej1988,Puche1991}. Here we apply two different methods to extract the rotation velocity of NGC~55 out to an unprecedented (for high-resolution, interferometric data) radius of almost $20~\mathrm{kpc}$:
  \begin{enumerate}
    \item the fitting of tilted rings to the radial velocity field;
    \item the envelope-tracing method operating on the po\-si\-tion--velocity diagram.
  \end{enumerate}
  Using two different methods will help us to better understand the uncertainties and systematic errors involved and obtain a more reliable measurement of the overall rotation curve despite the galaxy's unfavourable inclination. In the following, both methods will be briefly introduced and their results discussed and compared.
  
  \subsection{Tilted-ring model}
  
  The tilted-ring approach (e.g., \citealt{Rogstad1974}) attempts to fit a set of usually concentric tilted rings to the radial velocity field of a galaxy in order to extract the rotation velocity, inclination, and position angle of each ring under the assumption that the gas particles move on circular orbits at a constant speed. The method usually works best on galaxies with intermediate inclination angles and becomes increasingly difficult for galaxies that are almost face-on or edge-on.
  
  We followed the same approach as described in detail in Paper~1 for NGC~300, using the \textsc{gipsy} task \textsc{rotcur} \citep{Allen1985,Begeman1987,vanderHulst1992} for the tilted-ring fit. The only difference is that we used \textsc{rotcur} to determine the dynamical centre of NGC~55, because the galaxy does not feature a bright, point-like, optical nucleus. The mean position offset of the rings with respect to the optical centre of NGC~55 as listed in Table~\ref{tab_ngc55} is $\Delta \alpha = 3.4 \pm 1.0~\mathrm{arcmin}$ and $\Delta \delta = {-1.3} \pm 0.3~\mathrm{arcmin}$, implying that the dynamical centre of the galaxy is located about $3.5~\mathrm{arcmin}$ to the south-east of the optical centre, the latter of which is the position of the brightest pixel in the 2MASS J-band image \citep{Jarrett2000}.
  
\begin{figure*}
    \centering
    \includegraphics[width=0.8\linewidth]{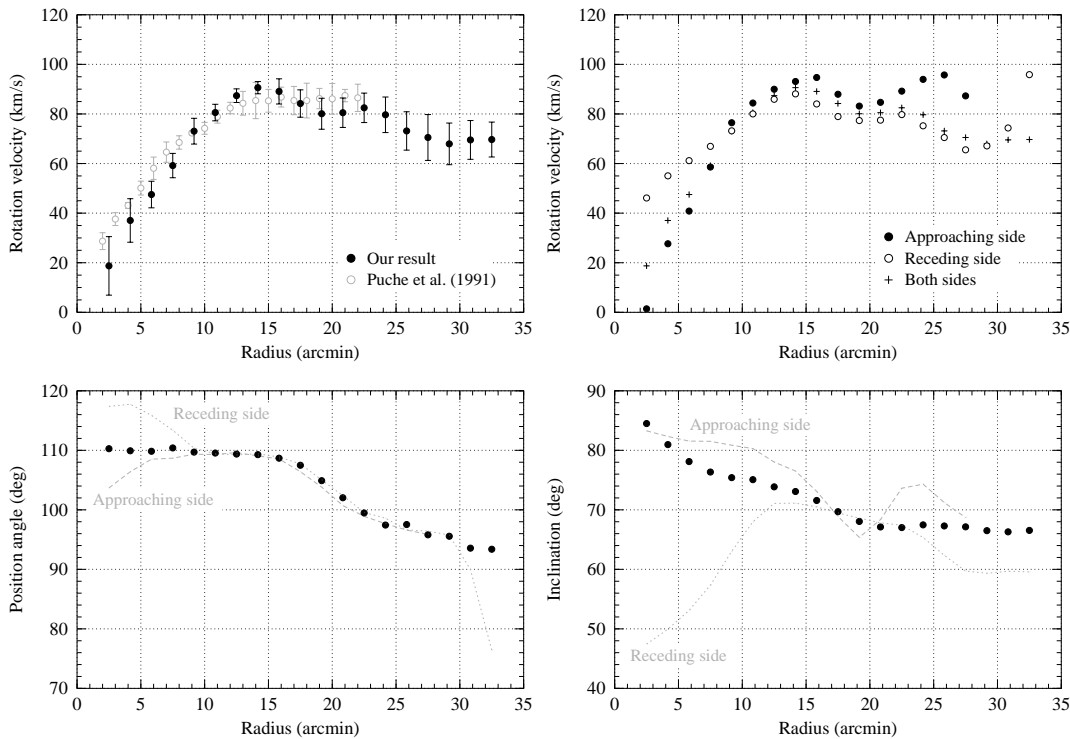}
    \caption{Results of the tilted-ring fit. \textit{Top-left panel:} Overall rotation curve of NGC~55 out to a radius of about $18~\mathrm{kpc}$. The rotation curve determined by \citet{Puche1991} based on VLA data is shown for comparison. \textit{Top-right panel:} Rotation curves derived from the approaching and receding side of NGC~55 separately compared to the overall rotation curve. \textit{Bottom-left and bottom-right panels:} Position angle, $\varphi$, and inclination angle, $i$, respectively, of the tilted-ring model used for the final rotation curve fit with \textsc{rotcur}. The dashed and dotted grey lines show the values derived separately for the approaching and receding side, respectively.}
    \label{fig_rotcur}
  \end{figure*}

  The results of the tilted-ring fit are shown in Fig.~\ref{fig_rotcur} and Table~\ref{tab_rotcur}. The top-left panel displays the final rotation curve of NGC~55, the error bars corresponding to the standard deviation about the mean rotation velocity along each ring. Separate fits for the approaching and receding sides of the galaxy are shown in the top-right panel of Fig.~\ref{fig_rotcur}.
  
  The rotation curve rises to a maximum of about $90~\mathrm{km \, s}^{-1}$ at a radius of $15~\mathrm{arcmin}$ or $8~\mathrm{kpc}$ and then gradually declines to about $70~\mathrm{km \, s}^{-1}$ at the outermost radius of $32.5~\mathrm{arcmin}$ or $18~\mathrm{kpc}$. Assuming a spherically-symmetric mass distribution, the total mass enclosed within the outermost data point is $M_{\rm tot} = (2.0 \pm 0.4) \times 10^{10}~\mathrm{M}_{\sun}$ which is almost the same as that of NGC~300 of $(2.9 \pm 0.2) \times 10^{10}~\mathrm{M}_{\sun}$ within a radius of $18.4~\mathrm{kpc}$ (Paper~1). The \ion{H}{i}~mass fraction of NGC~55 is $M_{\rm H\,I} / M_{\rm tot} \approx 8.5$~per cent. Our observations are much more sensitive in terms of surface brightness sensitivity than those of \citet{Puche1991}, allowing us to trace the rotation curve out to a much larger radius of $18~\mathrm{kpc}$, about 50~per cent farther out than the rotation curve of \citet{Puche1991}. It appears that this increased surface brightness sensitivity is vital in being able to detect the declining part of the rotation curve. Overall, the rotation curve of NGC~55 is similar to that of NGC~300 (Paper~1) which has a similar maximum and also declines across the outer parts of the disc.
  
  Position angle and inclination of the disc are plotted in the bottom-left and bottom-right panels of Fig.~\ref{fig_rotcur}, respectively. Again, the separate solutions for the approaching and receding halves of the disc are shown as well. Across the inner part of NGC~55 the position angle is constant at about $110\degr{}$ with respect to the J2000 equatorial coordinate system, while across the outer regions there is a moderate change to just under $95\degr{}$ near the outer edge of the detectable disc. The inclination angle gradually decreases from about $85\degr{}$ to just under $70\degr{}$ across the inner $20~\mathrm{arcmin}$ ($11~\mathrm{kpc}$) and then remains constant at about $67\degr{}$ across the outer parts of the disc of NGC~55.
  
\begin{figure}
    \centering
    \includegraphics[width=0.85\linewidth]{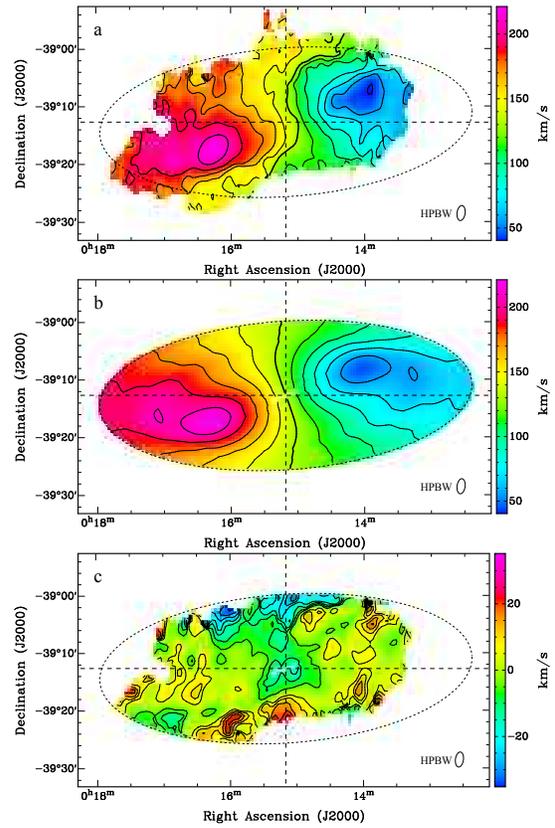}
    \caption{Tilted-ring model of NGC~55. \textbf{(a)}~Radial velocity field (barycentric) based on the fitting of Gauss--Hermite polynomials to the \ion{H}{i}~spectra. \textbf{(b)}~Radial velocity field derived from the tilted-ring model fit. The contours in this and the previous map are separated by $15~\mathrm{km \, s}^{-1}$ centred on the systemic velocity of NGC~55 (bold contour). \textbf{(c)}~Difference between the model and the observations. The contours cover the range of $\pm 35~\mathrm{km \, s}^{-1}$ in steps of $5~\mathrm{km \, s}^{-1}$, excluding $0$. In all three maps the dotted ellipse marks the edge of the outermost tilted ring used in the fit, and the dashed lines pass through the dynamical centre of NGC~55 and have been added to guide the eye.}
    \label{fig_velfit}
  \end{figure}

  To asses the quality of the tilted-ring fit, we compare in Fig.~\ref{fig_velfit} the observed velocity field of NGC~55 (based on fitting Gauss-Hermite polynomials to the \ion{H}{i}~spectra) with the velocity field resulting from the tilted-ring fit. While in general the model is able to adequately reproduce the velocity field, it fails to account for several of the intricate details seen in the observations.
  
  To begin with, the aforementioned global asymmetry in the \ion{H}{i}~disc of NGC~55 has strong effects on the velocity field, resulting in a larger separation between velocity contours on the receding side of NGC~55 than on the approaching side (see Fig.~\ref{fig_velfit}\,a). Secondly, the dynamical centre of NGC~55 does not coincide with either the optical centre or the point along the major axis where the velocity is equal to the systemic velocity of the galaxy. A tilted-ring model with concentric rings can in principle not reproduce these asymmetries, and Fig.~\ref{fig_velfit}\,c reveals substantial discrepancies in excess of $10~\mathrm{km \, s}^{-1}$ across the inner parts of NGC~55. The model also fails to explain the velocity of gas far away from the major axis. The strongly curved velocity contours near the far ends of the minor axis suggest a strong change in position angle for the outer rings. This is in conflict with the more regular appearance of the velocity field along the major axis which suggests only a moderate position angle change. Hence, some of the gas seen at larger separations from the major axis may be located beyond the disc and the result of infall, outflow, or some other form of distortion.
  
  In summary, the tilted-ring model does not describe the kinematics of NGC~55 very well and naturally fails to explain the major asymmetries in the \ion{H}{i}~disc. The model suggests that most of the gas at larger distances from the major axis is the result of warping of the \ion{H}{i}~disc (similar to NGC~300, see Paper~1) with gradual changes of both position angle and inclination across the disc, while some of the gas could also be located outside the disc. The tilted-ring model also seems to suggest that the rotation velocity decreases again in the outer disc of NGC~55.
  
  \begin{table}
    \centering
    \caption{Angular radius, $\vartheta$, physical radius, $R$, rotation velocity, $v_{\rm rot}$, position angle, $\varphi$, and inclination, $i$, of the tilted-ring model fitted to NGC~55.}
    \label{tab_rotcur}
    \begin{tabular}{rrrrr}
      \hline
       $\vartheta$ & $R$    & $v_{\rm rot}$   & $\varphi$ & $i$ \\
      ($''$) & (kpc) & ($\mathrm{km \, s}^{-1}$) & (deg) & (deg) \\
      \hline
       150 & $ 1.4$ & $18.7 \pm 11.8$ & $110.3$ & $84.5$ \\
       250 & $ 2.3$ & $37.0 \pm  8.8$ & $109.9$ & $81.0$ \\
       350 & $ 3.2$ & $47.5 \pm  5.4$ & $109.8$ & $78.1$ \\
       450 & $ 4.1$ & $59.2 \pm  4.9$ & $110.4$ & $76.4$ \\
       550 & $ 5.1$ & $73.1 \pm  5.2$ & $109.7$ & $75.4$ \\
       650 & $ 6.0$ & $80.6 \pm  3.3$ & $109.5$ & $75.1$ \\
       750 & $ 6.9$ & $87.4 \pm  2.8$ & $109.4$ & $73.9$ \\
       850 & $ 7.8$ & $90.6 \pm  2.5$ & $109.3$ & $73.1$ \\
       950 & $ 8.8$ & $89.1 \pm  5.1$ & $108.7$ & $71.6$ \\
      1050 & $ 9.7$ & $84.2 \pm  5.5$ & $107.5$ & $69.7$ \\
      1150 & $10.6$ & $80.1 \pm  6.3$ & $104.9$ & $68.1$ \\
      1250 & $11.5$ & $80.5 \pm  5.9$ & $102.0$ & $67.1$ \\
      1350 & $12.4$ & $82.5 \pm  5.9$ &  $99.5$ & $67.0$ \\
      1450 & $13.4$ & $79.7 \pm  7.2$ &  $97.4$ & $67.5$ \\
      1550 & $14.3$ & $73.1 \pm  7.7$ &  $97.5$ & $67.3$ \\
      1650 & $15.2$ & $70.5 \pm  9.3$ &  $95.8$ & $67.1$ \\
      1750 & $16.1$ & $67.9 \pm  8.4$ &  $95.6$ & $66.5$ \\
      1850 & $17.0$ & $69.5 \pm  7.8$ &  $93.6$ & $66.3$ \\
      1950 & $18.0$ & $69.7 \pm  7.0$ &  $93.4$ & $66.5$ \\

      \hline
    \end{tabular}
  \end{table}
  
  \subsection{Envelope-tracing method}
  
  The envelope-tracing method (e.g., \citealt{Sancisi1979,Sofue2001}) attempts to determine the rotation velocity at each position along the major axis of a galaxy by extracting the so-called `terminal velocity', $v_{\rm t}$, from the outer edge of the spectral line, i.e.\ the edge facing away from the systemic velocity of the galaxy. The method is usually applied to (almost) edge-on galaxies where the fitting of tilted rings becomes increasingly difficult.
  
  We followed the approach described by \citet{Sofue2001} to calculate the terminal velocity along the major axis of NGC~55 under the assumption that the gas is optically thin (which may not be the case in the innermost regions of NGC~55) and that the orbits within the disc are circular. We first extracted \ion{H}{i}~spectra from the position--velocity diagram along the major axis in \textsc{karma} \citep{Gooch1996}. Next, we calculated for each spectrum the intensity at the terminal velocity as defined by
  \begin{equation}
    I_{\rm t} = \sqrt{(\eta I_{\rm max})^{2} + I_{3 \sigma}^{2}}
  \end{equation}
  where $I_{\rm max}$ is the peak intensity of the spectral line, $I_{3 \sigma} = 6~\mathrm{mJy}$ corresponds to three times the noise level in the interpolated position--velocity diagram along the major axis, and $\eta$ was chosen to be $0.2$, thus selecting about 20~per cent of the peak intensity for spectra of high signal-to-noise ratio and approximately three times the noise level for spectra of low signal-to-noise ratio. The terminal velocity, $v_{\rm t}$, was then derived by progressing along the spectrum from the position of peak intensity outwards (i.e.~away from the systemic velocity) until the intensity was found to drop below $I_{\rm t}$. The terminal velocity was then extracted by linearly interpolating across the two bracketing channels in between which the drop below $I_{\rm t}$ occurred. The resulting terminal velocities are plotted as the red symbols (circles) on top of the position--velocity dia\-gram in Fig.~\ref{fig_posivelo}.
  
  With the terminal velocity, $v_{\rm t}$, known, we then calculated the rotation velocity at each position along the major axis via
  \begin{equation}
    v_{\rm rot} = \frac{|v_{\rm t} - v_{\rm sys}|}{\sin i} - \sqrt{\sigma_{\rm ISM}^{2} + \sigma_{\rm instr}^{2}}
  \end{equation}
  where $v_{\rm sys}$ is the systemic velocity of NGC~55 as listed in Table~\ref{tab_ngc55}, $i$ is the inclination of the gas disc, $\sigma_{\rm ISM}$ is the intrinsic velocity dispersion of the ISM, and $\sigma_{\rm instr}$ denotes the instrumental effect caused by the finite velocity resolution of the spectra. We assumed a constant inclination of $i = 80\degr{}$ across the entire disc \citep{KiszkurnoKoziej1988}, a velocity dispersion of the ISM of $\sigma_{\rm ISM} = 10~\mathrm{km \, s}^{-1}$ in line with \citet{Gentile2004}, and an instrumental effect of $\sigma_{\rm instr} = \Delta v / \sqrt{8 \ln 2}$, where $\Delta v = 8~\mathrm{km \, s}^{-1}$ is the spectral channel width of our data.
  
  The resulting rotation curve based on the envelope-tracing method is shown in Fig.~\ref{fig_envtrac} and Table~\ref{tab_envtrac} for the approaching and receding sides of NGC~55. Across the inner $10$ to $15~\mathrm{arcmin}$ the rotation velocity rises to a maximum of just under $100~\mathrm{km \, s}^{-1}$ and then gradually decreases again in the outer regions of the disc. A comparison between the rotation curves from the approaching and receding sides reveals the strong asymmetries in the \ion{H}{i}~disc of NGC~55 discussed earlier. On the approaching side, the rotation velocity rises much faster than on the receding side and then flattens off much earlier at a radius of about $8~\mathrm{arcmin}$. This asymmetry reflects the conspicuous kink in the position--velocity diagram that is clearly visible in Fig.~\ref{fig_posivelo}, confirming that the gas disc of NGC~55 is distorted.
  
  The total mass derived from the rotation velocity within a radius of $30.5~\mathrm{arcmin}$ or $17~\mathrm{kpc}$ (excluding the outermost data point seen in Fig.~\ref{fig_envtrac}) is $M_{\rm tot} = 1.7 \times 10^{10}~\mathrm{M}_{\sun}$, again assuming a spherically-symmetric mass distribution. This result is fully consistent with the mass of $(2.0 \pm 0.4) \times 10^{10}~\mathrm{M}_{\sun}$ derived from the tilted-ring model at a radius of $18~\mathrm{kpc}$.
  
  \subsection{Comparison}
  
  A comparison between the rotation curves derived from the tilted-ring model and the envelope-tracing method (Fig.~\ref{fig_envtrac}) reveals that they agree well in the outer regions of NGC~55 beyond about $15$ to $20~\mathrm{arcmin}$ ($8$ to $11~\mathrm{kpc}$), where both rotation curves have approximately the same amplitude and show a gradual decline with a similar slope. In the inner regions, however, there is a strong discrepancy, with the tilted-ring rotation velocities being systematically smaller than those derived from the envelope-tracing method.
  
  This result, however, is not unexpected, given the high inclination of most of the disc of NGC~55. In an almost edge-on gas disc, a line of sight passing through the disc close to the galaxy's centre will cover a large range of azimuthal angles. Hence, the peak in the integrated spectrum along that line of sight will be systematically shifted away from the terminal velocity and towards the systemic velocity of the galaxy, resulting in a measured rotation velocity that is too small. This effect becomes less pronounced in the outer regions of the disc, where the azimuthal range covered by a line of sight is smaller. In the outer parts of the disc of NGC~55 our rotation curve from the tilted-ring model is therefore more accurate and in good agreement with that from the envelope-tracing method. However, the tilted-ring rotation curve may not be reliable and should not be trusted at smaller radii of $\vartheta \la 15~\mathrm{arcmin}$.
  
  For the envelope-tracing method we assumed a constant inclination of $80\degr{}$ across the entire galaxy, whereas the tilted-ring model yields a moderate decrease in the inclination angle from about $80\degr{}$ near the centre to approximately $67\degr{}$ across the outer disc of NGC~55. If genuine, this decrease in the inclination angle would result in a moderate increase of the rotation velocity derived from the envelope-tracing method of about 7~per cent. After correction for the velocity dispersion of the gas and the instrumental resolution, this would result in only a small increase in the rotation velocity from about $70$ to $75~\mathrm{km \, s}^{-1}$ in the outer regions of NGC~55. Hence, the rotation curve derived from the envelope-tracing method would still be consistent with the tilted-ring model in the outer parts of the galaxy, thereby substantiating the observed decline in the rotation velocity of NGC~55 beyond a radius of about $15~\mathrm{arcmin}$ or $8~\mathrm{kpc}$.

  \section{Discussion}
  \label{sect_discussion}
  
  \subsection{Origin of the distortions in the gas disc of NGC~55}
  
\begin{figure}
    \centering
    \includegraphics[width=\linewidth]{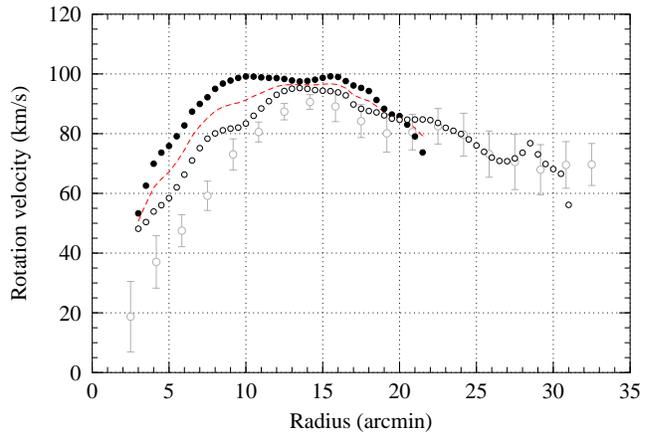}
    \caption{Rotation curve of NGC~55 as determined by the envelope-tracing method along the major axis for the approaching (filled black circles) and receding (open black circles) sides of the galaxy. The dashed red line shows the average of the two. The rotation curve determined from the tilted-ring model (open grey circles) is shown for comparison.}
    \label{fig_envtrac}
  \end{figure}

  As discussed in Sections~\ref{sect_ngc55} and~\ref{sect_outerdisc}, the inner and outer regions of the vast \ion{H}{i}~disc of NGC~55 show signs of strong distortions. One of the most intriguing distortions is the conspicuous kink in the position--velocity diagram along the major axis of the disc as presented in Fig.~\ref{fig_posivelo}. As a result of this kink, the rotation velocity across the western, approaching half of NGC~55 appears to be rising much faster than in the eastern, receding part of the galaxy. This marked asymmetry is also obvious from the rotation curves in Fig.~\ref{fig_envtrac} as derived from the envelope-tracing method.
  
  Such a strong kinematic asymmetry on a global scale must be connected to underlying asymmetries in the overall structure of NGC~55, and the most obvious connection in this case would be the bar that dominates optical images of the galaxy. The orientation of the bar along the line of sight means that we are seeing the bar almost end-on. Hence, any non-circular motion of gas along the bar axis would result in a strong radial velocity gradient across a relatively small angular distance on the sky along the projected length of the bar, similar to what we observe in the position--velocity diagram across the western half of the galaxy.
  
  The effect was modelled by \citet{Rodriguez-Fernandez2008} for the Milky Way, and their figure~11 serves to illustrate the effect of an increase in the radial velocity range across the bar for a decreasing angle between the bar and the line of sight. The situation in the Milky Way is different from that in NGC~55, though, where the bar dominates the overall structure of the galaxy. This might explain the relative dominance of the kinematic effect seen in NGC~55.
  
  While the asymmetry seen in the inner disc can be understood as a consequence of the presence of a bar in NGC~55, the complex structure and distortions of the outer disc are more difficult to explain. First of all, there is a strong global asymmetry in the column density profiles along the major and minor axis of NGC~55, as depicted in Fig.~\ref{fig_profiles}. In the northern and eastern part the disc is significantly more extended and forms a noticeable `tail' at the eastern end of the disc, whereas the disc is less extended in the south and west, where it ends quite abruptly.
  
  \begin{table}
    \centering
    \caption{Angular radius, $\vartheta$, physical radius, $R$, and rotation velocity, $v_{\rm rot}$, for the approaching and receding side of NGC~55 as derived from the envelope-tracing method.}
    \label{tab_envtrac}
    \begin{tabular}{rrrrrrrr}
      \hline
      $\vartheta$ & $R$ & $v_{\rm rot}^{\rm app}$ & $v_{\rm rot}^{\rm rec}$ & $\vartheta$ & $R$ & $v_{\rm rot}^{\rm app}$ & $v_{\rm rot}^{\rm rec}$ \\
      ($'$) & (kpc) & \multicolumn{2}{c}{($\mathrm{km \, s}^{-1}$)} & ($'$) & (kpc) & \multicolumn{2}{c}{($\mathrm{km \, s}^{-1}$)} \\
      \hline
      $ 3.0$ & $1.66$ & $53.3$ & $48.2$ & $17.5$ & $ 9.67$ & $95.3$ & $88.3$ \\
      $ 3.5$ & $1.93$ & $62.6$ & $50.4$ & $18.0$ & $ 9.95$ & $94.3$ & $87.6$ \\
      $ 4.0$ & $2.21$ & $69.9$ & $53.9$ & $18.5$ & $10.22$ & $91.3$ & $87.1$ \\
      $ 4.5$ & $2.49$ & $73.6$ & $56.1$ & $19.0$ & $10.50$ & $88.3$ & $86.1$ \\
      $ 5.0$ & $2.76$ & $75.9$ & $58.4$ & $19.5$ & $10.78$ & $86.4$ & $85.0$ \\
      $ 5.5$ & $3.04$ & $79.1$ & $62.0$ & $20.0$ & $11.05$ & $85.9$ & $84.7$ \\
      $ 6.0$ & $3.32$ & $82.7$ & $66.3$ & $20.5$ & $11.33$ & $83.0$ & $84.7$ \\
      $ 6.5$ & $3.59$ & $87.4$ & $71.0$ & $21.0$ & $11.61$ & $79.1$ & $84.7$ \\
      $ 7.0$ & $3.87$ & $90.0$ & $75.2$ & $21.5$ & $11.88$ & $73.7$ & $84.8$ \\
      $ 7.5$ & $4.15$ & $92.2$ & $78.3$ & $22.0$ & $12.16$ &     -- & $84.5$ \\
      $ 8.0$ & $4.42$ & $95.0$ & $80.1$ & $22.5$ & $12.44$ &     -- & $83.6$ \\
      $ 8.5$ & $4.70$ & $96.8$ & $81.1$ & $23.0$ & $12.71$ &     -- & $81.9$ \\
      $ 9.0$ & $4.97$ & $97.8$ & $81.7$ & $23.5$ & $12.99$ &     -- & $81.0$ \\
      $ 9.5$ & $5.25$ & $98.7$ & $82.0$ & $24.0$ & $13.26$ &     -- & $79.9$ \\
      $10.0$ & $5.53$ & $99.2$ & $83.4$ & $24.5$ & $13.54$ &     -- & $78.1$ \\
      $10.5$ & $5.80$ & $99.1$ & $86.0$ & $25.0$ & $13.82$ &     -- & $76.1$ \\
      $11.0$ & $6.08$ & $98.8$ & $88.4$ & $25.5$ & $14.09$ &     -- & $73.9$ \\
      $11.5$ & $6.36$ & $98.7$ & $90.9$ & $26.0$ & $14.37$ &     -- & $72.0$ \\
      $12.0$ & $6.63$ & $98.6$ & $93.0$ & $26.5$ & $14.65$ &     -- & $70.9$ \\
      $12.5$ & $6.91$ & $98.3$ & $94.3$ & $27.0$ & $14.92$ &     -- & $70.8$ \\
      $13.0$ & $7.18$ & $97.8$ & $95.1$ & $27.5$ & $15.20$ &     -- & $71.7$ \\
      $13.5$ & $7.46$ & $97.5$ & $95.3$ & $28.0$ & $15.48$ &     -- & $73.6$ \\
      $14.0$ & $7.74$ & $97.7$ & $95.0$ & $28.5$ & $15.75$ &     -- & $76.8$ \\
      $14.5$ & $8.01$ & $98.1$ & $94.6$ & $29.0$ & $16.03$ &     -- & $73.1$ \\
      $15.0$ & $8.29$ & $98.7$ & $94.4$ & $29.5$ & $16.30$ &     -- & $69.8$ \\
      $15.5$ & $8.57$ & $99.2$ & $94.2$ & $30.0$ & $16.58$ &     -- & $68.1$ \\
      $16.0$ & $8.84$ & $99.0$ & $93.8$ & $30.5$ & $16.86$ &     -- & $66.6$ \\
      $16.5$ & $9.12$ & $97.6$ & $92.8$ & $31.0$ & $17.13$ &     -- & $56.2$ \\
      $17.0$ & $9.40$ & $96.1$ & $89.8$ &        &         &        &        \\
      \hline
    \end{tabular}
  \end{table}
  
  This asymmetry is remarkably similar to the one seen in neighbouring NGC~300, which we attributed to ram-pressure interaction with the intergalactic medium in the Sculptor group (see Paper~1). A basic calculation revealed that under reasonable assumptions on the density of the IGM, $n \simeq 10^{-5}$ to $10^{-4}~\mathrm{cm}^{-3}$, and the relative velocity of NGC~300, $v \simeq 200~\mathrm{km \, s}^{-1}$, ram-pressure forces have the potential to distort the \ion{H}{i}~gas near the outer edge of the disc of NGC~300. These results would apply to NGC~55 as well under the assumption that both galaxies have a comparable mass and experience similar IGM densities, as expected from their proximity and common group membership. Hence, NGC~55 would be affected by ram pressure in the same way as NGC~300, resulting in similar distortions of the outer \ion{H}{i}~disc of NGC~55, in agreement with our observational results.
  
\begin{figure}
    \centering
    \includegraphics[width=\linewidth]{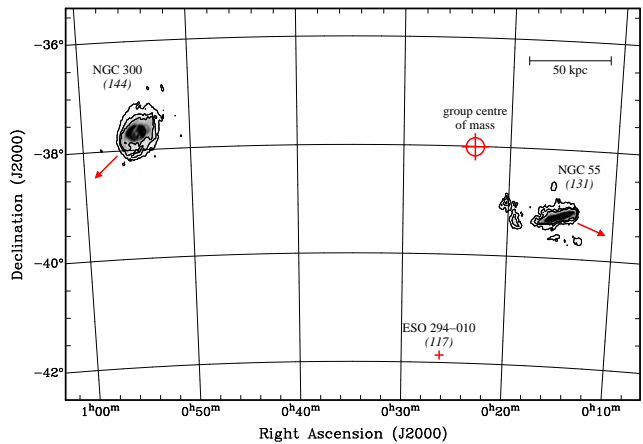}
    \caption{Combined \ion{H}{i} column density map of NGC~55 and 300, showing both galaxies at the correct position and scale. The contour levels are $0.2$, $1.0$, and $2.0 \times 10^{20}~\mathrm{cm}^{-2}$. The red arrows indicate the approximate direction of motion relative to the IGM in the Sculptor group under the assumption that the characteristic asymmetries in the outer gas discs of the two galaxies are caused by ram-pressure interaction with the IGM. The red cross marks the position of the dwarf galaxy ESO~294$-$010. The numbers in brackets denote barycentric velocities in $\mathrm{km \, s}^{-1}$. The red cross-hair marks the centre of mass of the NGC~55 sub-group according to \citet{Garcia1993}.}
    \label{fig_sculptor}
  \end{figure}

  If the global asymmetries in the outer \ion{H}{i}~discs of NGC~55 and~300 were indeed caused by ram-pressure interaction, we could use the orientation of their `tails' to deduce the approximate projected direction of motion of the two galaxies relative to the IGM of the Sculptor group. In this case, NGC~300 would be moving in a south-easterly direction (with respect to the IGM, not the celestial coordinate system), while NGC~55 would be moving towards the south-west at a slightly smaller angle with respect to the right ascension axis (Fig.~\ref{fig_sculptor}). For reference, we also show in Fig.~\ref{fig_sculptor} the position of the centre of mass of the NGC~55 sub-group (consisting of NGC~55, NGC~300, NGC~7793, and four dwarf galaxies) according to the Lyon Groups of Galaxies catalogue \citep{Garcia1993}. This geometry suggests that NGC~55 and~300 were much closer to one another in the past and are currently moving away from each other, raising the question of whether there could have been a close encounter between the two galaxies in the past that could have led to the observed distortions and asymmetries in the outer gas discs of both galaxies.
  
  The possibility of a recent encounter with either NGC~300 or another Sculptor group member was already mentioned by \citet{Hummel1986} and \citet{Puche1991} based on the observed asymmetries in both the velocity field and the distribution of the \ion{H}{i}~gas. Recent distance estimates place both NGC~55 and NGC~300 at approximately the same distance of $d = 1.9~\mathrm{Mpc}$ \citep{Gieren2004,Pietrzynski2006}. This implies that the distance, $D$, between the two galaxies can directly be inferred from their angular separation on the sky of $\varphi = 8\degr{}$, hence $D = d \, \tan(\varphi) \approx 270~\mathrm{kpc}$, which is merely eight times the diameter of the \ion{H}{i}~disc of NGC~300 of about $35~\mathrm{kpc}$. Consequently, NGC~55 and~300 form a relatively close pair (Fig.~\ref{fig_sculptor}), their separation being only about a third of the distance between the Andromeda Galaxy and the Milky Way (e.g.~\citealt{Vilardell2007}).
  
  At the same time, both galaxies also have approximately the same radial velocity of $131$ and $144~\mathrm{km \, s}^{-1}$ in the barycentric reference frame for NGC~55 and 300, respectively. Assuming a significant relative velocity of NGC~55 with respect to NGC~300, this implies that the vector of the relative velocity between the two galaxies must lie almost exactly in the plane of the sky. An object travelling at a constant velocity of $200~\mathrm{km \, s}^{-1}$ would be able to traverse the current distance between NGC~55 and NGC~300 in approximately $1.3~\mathrm{Ga}$. This result, although based on a greatly simplified estimate, illustrates that NGC~55 and 300 could well have been much closer to each other at some point during the past one or two billion years, and it would seem likely that such an encounter could have resulted in the strong distortions and warping observed in the \ion{H}{i}~discs of both galaxies today.
  
  If we assume a rotation velocity of $v_{\rm rot} = 70~\mathrm{km \, s}^{-1}$ at the outermost point of the rotation curve of $r = 18~\mathrm{kpc}$, the resulting orbital period for gas particles near the edge of the \ion{H}{i}~disc is about $1.6~\mathrm{Ga}$. Hence, the gas in outer disc of NGC~55 has only completed about one orbit since the possible encounter with NGC~300. This may not be enough for the outer disc of NGC~55 to dynamically settle from potential disturbances induced by the gravitational pull of NGC~300 -- and vice versa -- supporting the concept that the substantial warping observed in the discs of the two galaxies today could have originated from a previous close encounter of NGC~55 and~300 between $1$ and $2~\mathrm{Ga}$ ago.
  
  An alternative mechanism to explain the warping seen in NGC~55 could be a recent interaction with a smaller satellite galaxy. The nearest dwarf galaxy to NGC~55 currently known is ESO~294$-$010 at a barycentric radial velocity of $117~\mathrm{km \, s}^{-1}$ \citep{Jerjen1998} and an angular separation of $3.5~\mathrm{degrees}$, corresponding to a projected distance of only about $115~\mathrm{kpc}$. Another option would be that the satellite responsible for the distortions has already been accreted on to the disc of NGC~55, thereby generating the numerous extra-planar gas features visible in Fig.~\ref{fig_disc}.
  
  The assumption of a recent interaction with a satellite galaxy is supported by deep optical photometry of a small section covering the northern part of the stellar disc of NGC~55 by \citet{Tanaka2011}. By selecting populations of red and blue RGB stars through simple colour and metallicity cuts they reveal the presence of a thick disc in NGC~55 with a scale height of $z \approx 1.6~\mathrm{kpc}$. The thick disc shows a marked asymmetry, the eastern part having a significantly larger vertical extent than the western part. \citet{Tanaka2011} interpret this asymmetry as the possible signature of an accretion event affecting the disturbed eastern part of the disc. The increased scale height of the thick disc in the eastern part coincides with an increased thickness of the \ion{H}{i}~disc in that region (near substructure~2 in Fig.~\ref{fig_disc}). Distortions in the stellar disc can only occur as the result of gravitational forces, and it seems likely that the interaction or merging event that was capable of creating the noticeable asymmetry of the thick disc is also responsible for some of the distortions of the \ion{H}{i}~disc in the eastern half of NGC~55, potentially forming some of the spurs and clumps observed in the outer regions of the disc.
  
  \subsection{Origin of the isolated clouds}
  
  The origin of some of the brighter and more extended clouds seen in the vicinity of NGC~55 and~300 \citep{Mathewson1975,Haynes1979} has been controversial, resulting in vigorous discussions in the literature. \citet{Mathewson1975} speculated that the clouds, and other high-velocity clouds (HVCs) found in the northern sky by other observers, were forming an intergalactic population, with the clouds near NGC~55 and~300 being located within the Sculptor group. That opinion was refuted by \citet{Haynes1979} based on the clouds' positions and velocities. Instead, \citet{Haynes1979} argued that the clouds were more likely to be part of the Magellanic Stream which runs across the same region on the sky. Their notion was supported by \citet{Putman2003} who mapped the entire Magellanic Stream in \ion{H}{i} with the 64-m Parkes telescope and found no convincing evidence for an association of the clouds with the Sculptor group.
  
  As discussed in detail in Section~\ref{sect_hvcs}, there is evidence for the brightest and most extended clouds detected in our data, clouds~6 and~7, to be foreground objects and part of the presumed population of Magellanic Stream clouds studied by \citet{Haynes1979} and \citet{Putman2003}. The remaining clouds, however, are conspicuously concentrated near NGC~55, with small angular separations from the disc and radial velocities similar to those of NGC~55, leading us to conclude that they are distinct from the population described by \citet{Haynes1979} and \citet{Putman2003} and instead form a circum-galactic population surrounding NGC~55.
  
  Populations of circum-galactic gas clouds or HVCs are not uncommon and have been discovered around several other external galaxies, most notably the Andromeda Galaxy \citep{Thilker2004}, M33 \citep{Putman2009}, the M81~group \citep{Chynoweth2008,Chynoweth2011}, and M83 \citep{Miller2009}.
  
  The population of HVCs around M31, discovered by \citet{Thilker2004} and studied in detail by \citet{Westmeier2005} and \citet{Westmeier2008}, is located within about $50~\mathrm{kpc}$ of M31. Possible origins of the clouds include tidal debris from interactions and mergers, gaseous counterparts of dark-matter satellites, or accretion of cooling gas from the corona of M31 or the intergalactic medium of the Local Group. \citet{Chynoweth2008} conclude that the clouds found in the M81~group are likely the result of tidal interaction between the galaxies in the group, while \citet{Miller2009} assume a combination of tidal stripping and galactic fountain \citep{Shapiro1976} for the HVCs near M83. Unfortunately, the information gathered from \ion{H}{i} emission-line observations alone is seldom sufficient to constrain the origin of gas clouds in the vicinity of galaxies, and our observations are no exception in this regard. In the case of NGC~55 the proximity (in projection) of the clouds to the disc of the galaxy speaks against their being primordial, gaseous satellites. At such close distances the clouds would be subject to strong ram-pressure forces exerted by the halo and disc gas of NGC~55 and would have rather short lifetimes inconsistent with the concept of a stable orbit.
  
  A more promising scenario could be tidal stripping. As discussed earlier, tidal interaction and accretion of satellites could be responsible for some of the distortions of the \ion{H}{i}~disc of NGC~55, and there is evidence that cloud number~5 is related to the southern spur of the disc. It seems likely that tidal interaction and accretion events capable of producing the strong distortions observed in the \ion{H}{i}~disc of the galaxy could also have created some of the isolated gas clouds. This would naturally explain the small projected distances of most clouds from the disc.
  
  It is also remarkable that almost all of the clouds have velocities close to the systemic velocity of NGC~55. Six of the eight clouds are within $\pm 45~\mathrm{km \, s}^{-1}$ of the systemic velocity, and four of those are even within $\pm 20~\mathrm{km \, s}^{-1}$. However, this observation cannot be considered as a strong argument for or against either tidal stripping or dark-matter satellites.
  
  The concentration of clouds near the disc of NGC~55 with velocities close to the galaxy's systemic velocity would also be consistent with the concept of gas being ejected from (or falling towards) the disc of NGC~55 by some internal mechanism. If we assume that the gas is being ejected approximately perpendicular to the disc plane, then its radial velocity, as seen by a terrestrial observer, could well be close to systemic due to the disc's high inclination of about $70$ to $80$~degrees. One such mechanism, which has been discussed extensively in the literature, is the so-called `galactic fountain' \citep{Shapiro1976}, by which ionised gas ejected from the disc by supernovae would be injected into the galactic corona, where it would cool down, condense high above the disc, and fall back towards the disc on a ballistic trajectory. The galactic fountain mechanism has been proposed to explain the existence of some of the high- and intermediate-velocity clouds of the Milky Way \citep{Bregman1980,Spitoni2008}, although others have questioned this hypothesis based on theoretical considerations \citep{Ferrara1992} and metallicity measurements of HVCs (\citealt{Binney2009}; also see \citealt{Wakker1997}).

  \section{Summary}
  
  We have mapped a large region of approximately $2\degr{} \! \times 2\degr{}$ around the Sculptor group galaxy NGC~55 in the 21-cm emission line of \ion{H}{i} with the ATCA. At a $5 \sigma$ column density sensitivity of $10^{19}~\mathrm{cm}^{-2}$ across a spectral channel width of $8~\mathrm{km \, s}^{-1}$ our observations reveal for the first time the vast extent of the galaxy's neutral gas disc at a moderately high spatial resolution on the order of $1~\mathrm{kpc}$. The main findings from our observations are:
  \begin{enumerate}
    \item The \ion{H}{i}~disc of NGC~55 has a diameter of just under $1\degr{}$, corresponding to about $31~\mathrm{kpc}$, within the $10^{19}~\mathrm{cm}^{-2}$ column density contour as measured along the major axis. The total \ion{H}{i}~mass of $(1.7 \pm 0.1) \times 10^{9}~\mathrm{M}_{\sun}$ as determined from our data is comparable to the mass detected in HIPASS \citep{Koribalski2004}, indicating that we are not missing much flux on large angular scales. Nevertheless, our flux measurement should be considered a lower limit due to optical depth effects and missing short spacings.
    \item Determination of the rotation curve through tilted-ring fitting and envelope tracing yields a maximum rotation velocity in the range of $90$ to $100~\mathrm{km \, s}^{-1}$. Beyond a galactocentric radius of about $15~\mathrm{arcmin}$ ($8~\mathrm{kpc}$) the rotation velocity gently decreases to about $70~\mathrm{km \, s}^{-1}$ near the outer edge of the disc. The resulting total mass within a radius of $18~\mathrm{kpc}$ is $(2.0 \pm 0.4) \times 10^{10}~\mathrm{M}_{\sun}$, suggesting an \ion{H}{i}~mass fraction of just under 10~per cent of the total mass.
    \item The \ion{H}{i}~disc of NGC~55 shows a strong global asymmetry in the column density distribution, with a conspicuous `tail' in the east, but a sharp and abrupt `edge' in the west. This asymmetry is presumably caused by ram-pressure forces as the galaxy is moving through the intergalactic medium of the Sculptor group. In this case, NGC~55 would be moving in a south-westerly direction and away from NGC~300, suggesting that both galaxies could have been much closer to each other over the past $1$ or $2~\mathrm{Ga}$.
    \item There is a conspicuous kink in the position--velocity diagram along the major axis which can be interpreted as the kinematic signature of non-circular motions along the bar of NGC~55, which is seen end-on and dominates the western part of the galaxy in optical images.
    \item Several conspicuous clumps and spurs found all across the \ion{H}{i}~disc of NGC~55 suggest that external or internal processes, such as tidal stripping or gas outflows, are continuously stirring up and shaping the disc. We also find that the gas disc is significantly warped, presumably as the result of tidal interaction with NGC~300 or a satellite galaxy.
    \item NGC~55 is surrounded by a population of isolated \ion{H}{i}~clouds within about $20~\mathrm{kpc}$ projected distance of the disc, which could be the equivalent of the HVC population seen around the Milky Way. While there is no ultimate proof for any individual cloud to be located in the vicinity of NGC~55, the dynamical properties of the clouds as well as their spatial distribution suggest that most of them are associated with NGC~55, whereas two of the clouds are most likely part of the Magellanic Stream in the foreground.
    \item From our \ion{H}{i}~data alone we are not able to determine the origin of the clouds, but their physical properties and distribution suggest that they are the result of tidal stripping, accretion, or outflow from the disc rather than gaseous dark-matter haloes surrounding NGC~55.
  \end{enumerate}
  
  \section*{Acknowledgments}
  
  The Australia Telescope Compact Array is part of the Australia Telescope which is funded by the Commonwealth of Australia for operation as a National Facility managed by CSIRO.

  \appendix
  
  \section{Remarks on the results of Puche, Carignan \& Wainscoat~(1991)}
  \label{app_puche}
  
  Our measured barycentric radial velocity of NGC~55 of $v_{\rm sys} = 131 \pm 2~\mathrm{km \, s}^{-1}$ is inconsistent with the barycentric velocity of $v_{\rm sys} = 118.3 \pm 4.0~\mathrm{km \, s}^{-1}$ derived by \citet{Puche1991} from the first moment of their integrated \ion{H}{i}~spectrum. An investigation of this discrepancy revealed that the radial velocity scale used by \citet{Puche1991} is presumably wrong.
  
  As noted earlier, the velocity axis labels and designation of the south-eastern and north-western side of NGC~55 in figure~5 of \citet{Puche1991} are apparently wrong and inconsistent with our spectrum in Fig.~\ref{fig_intspec}. The velocity of $0~\mathrm{km \, s}^{-1}$ should be on the right-hand side of their spectrum. To investigate this matter further, we obtained a copy of their original, reduced VLA data cube from the NASA/IPAC Extragalactic Database,\footnote{The NASA/IPAC Extragalactic Database (NED) is operated by the Jet Propulsion Laboratory, California Institute of Technology, under contract with the National Aeronautics and Space Administration.} created an integrated \ion{H}{i}~spectrum, and computed the first moment of that spectrum. The generated spectrum has the correct velocity scale, and the resulting barycentric radial velocity of $130~\mathrm{km \, s}^{-1}$ is fully consistent with the velocity derived from our data and clearly disagrees with the value quoted by \citet{Puche1991} in their paper.
  
  This result suggests that the inconsistent axis labels in figure~5 of \citet{Puche1991} are presumably not just an error in the artwork. When we take the integrated spectrum from their VLA data cube and deliberately flip the velocity axis in the same way as shown in their figure~5 ($v' = {-v} + 250~\mathrm{km \, s}^{-1}$, such that $250~\mathrm{km \, s}^{-1}$ becomes $0~\mathrm{km \, s}^{-1}$ and vice versa) we get an intensity-weighted mean velocity of $118~\mathrm{km \, s}^{-1}$ which is identical to the value of $v_{\rm sys} = 118.3 \pm 4.0~\mathrm{km \, s}^{-1}$ quoted by \citet{Puche1991} in their paper. This leads us to conclude that the authors possibly used a flipped velocity scale in their data analysis, and some of the velocity measurements presented by \citet{Puche1991} are presumably wrong and affected by this error. It is interesting to note, though, that the velocity designations in their channel maps (figure~3) and position--velocity diagram (figure~9) are correct and not affected by the error.
  
  \bsp
  
  \label{lastpage}

\begin{thebibliography}{99}
    \bibitem[\protect\citeauthoryear{Allen, Ekers \& Terlouw}{1985}]{Allen1985}
    Allen R.~J., Ekers R.~D., Terlouw, J.~P., 1985, in Di~Ges{\`u} V., Scarsi L., Crane P., Friedman J.~H., Levialdi S., eds, Data Analysis in Astronomy, Plenum Press, New York, USA, p.~271
    \bibitem[\protect\citeauthoryear{Barnes et al.}{2001}]{Barnes2001}
    Barnes D.~G.\ et al., 2001, MNRAS, 322, 486
    \bibitem[\protect\citeauthoryear{Begeman}{1987}]{Begeman1987}
    Begeman, K.~G., 1987, \ion{H}{i}~rotation curves of spiral galaxies, Ph.~D.\ thesis, Rijksuniversiteit Groningen
    \bibitem[\protect\citeauthoryear{Binney, Nipoti \& Fraternali}{2009}]{Binney2009}
    Binney J., Nipoti C., Fraternali F., 2009, MNRAS, 397, 1804
    \bibitem[\protect\citeauthoryear{Bregman}{1980}]{Bregman1980}
    Bregman J.~N., 1980, ApJ, 236, 577
    \bibitem[\protect\citeauthoryear{Castro et al.}{2008}]{Castro2008}
    Castro N.\ et al., 2008, A\&A, 485, 41
    \bibitem[\protect\citeauthoryear{Chynoweth et al.}{2008}]{Chynoweth2008}
    Chynoweth K.~M., Langston G.~I., Yun M.~S., Lockman F.~J., Rubin K.~H.~R., Scoles S.~A., 2008, AJ, 135, 1983
    \bibitem[\protect\citeauthoryear{Chynoweth, Langston \& Holley-Bockelmann}{2011}]{Chynoweth2011}
    Chynoweth K.~M., Langston G.~I., Holley-Bockelmann K., 2011, AJ, 141, 9
    \bibitem[\protect\citeauthoryear{Clark}{1965}]{Clark1965}
    Clark, B.~G., 1965, ApJ, 142, 1398
    \bibitem[\protect\citeauthoryear{Dale et al.}{2009}]{Dale2009}
    Dale D.~A.\ et al., 2009, ApJ, 703, 517
    \bibitem[\protect\citeauthoryear{Davidge}{2005}]{Davidge2005}
    Davidge T.~J., 2005, ApJ, 622, 279
    \bibitem[\protect\citeauthoryear{de~Vaucouleurs}{1961}]{deVaucouleurs1961}
    de~Vaucouleurs G., 1961, ApJ, 133, 405
    \bibitem[\protect\citeauthoryear{de~Vaucouleurs et al.}{1991}]{deVaucouleurs1991}
    de~Vaucouleurs G., de~Vaucouleurs A., Corwin H.~G., Buta R.~J., Paturel G., Fouque P., 1991, Third Reference Catalogue of Bright Galaxies. Springer-Verlag, Berlin, Heidelberg, New York
    \bibitem[\protect\citeauthoryear{Engelbracht et al.}{2004}]{Engelbracht2004}
    Engelbracht C.~W.\ et al., 2004, ApJS, 154, 248
    \bibitem[\protect\citeauthoryear{Epstein}{1964}]{Epstein1964}
    Epstein E.~E., 1964, AJ, 69, 490
    \bibitem[\protect\citeauthoryear{Ferguson, Wyse \& Gallagher}{1996}]{Ferguson1996}
    Ferguson A.~M.~N., Wyse R.~F.~G., Gallagher J.~S., 1996, AJ, 112, 2567
    \bibitem[\protect\citeauthoryear{Ferrara \& Einaudi}{1992}]{Ferrara1992}
    Ferrara A., Einaudi G., 1992, ApJ, 395, 475
    \bibitem[\protect\citeauthoryear{Garcia}{1993}]{Garcia1993}
    Garcia, A.~M., 1993, A\&AS, 100, 47
    \bibitem[\protect\citeauthoryear{Gentile et al.}{2004}]{Gentile2004}
    Gentile G., Salucci P., Klein U., Vergani D., Kalberla P., 2004, MNRAS, 351, 903
    \bibitem[\protect\citeauthoryear{Gieren et al.}{2004}]{Gieren2004}
    Gieren W.\ et al., 2004, AJ, 128, 1167
    \bibitem[\protect\citeauthoryear{Gooch}{1996}]{Gooch1996}
    Gooch R., 1996, in Jacoby G.~H., Barnes J., eds, Astronomical Data Analysis Software and Systems~V, ASP Conference Series, vol.~101, p.~80
    \bibitem[\protect\citeauthoryear{Haynes \& Roberts}{1979}]{Haynes1979}
    Haynes M.~P., Roberts M.~S., 1979, ApJ, 227, 767
    \bibitem[\protect\citeauthoryear{Hummel, Dettmar \& Wielebinski}{1986}]{Hummel1986}
    Hummel E., Dettmar R.-J., Wielebinski R., 1986, A\&A, 166, 97
    \bibitem[\protect\citeauthoryear{Jarrett et al.}{2000}]{Jarrett2000}
    Jarrett T.~H., Chester T., Cutri R., Schneider S., Skrutskie M., Huchra J.~P., 2000, AJ, 119, 2498
    \bibitem[\protect\citeauthoryear{Jerjen, Freeman \& Binggeli}{1998}]{Jerjen1998}
    Jerjen H., Freeman K.~C., Binggeli B., 1998, AJ, 116, 2873
    \bibitem[\protect\citeauthoryear{Karachentsev et al.}{2003}]{Karachentsev2003}
    Karachentsev I.~D.\ et al., 2003, A\&A, 404, 93
    \bibitem[\protect\citeauthoryear{Kiszkurno-Koziej}{1988}]{KiszkurnoKoziej1988}
    Kiszkurno-Koziej E., 1988, A\&A, 196, 26
    \bibitem[\protect\citeauthoryear{Koribalski}{2010}]{Koribalski2010}
    Koribalski B.~S., 2010, in Verdes-Montenegro L., Del Olmo A., Sulentic J., eds, Galaxies in Isolation: Exploring Nature Versus Nurture. ASP Conf.\ Ser., 421, 137
    \bibitem[\protect\citeauthoryear{Koribalski et al.}{2004}]{Koribalski2004}
    Koribalski B.~S.\ et al., 2004, AJ, 128, 16
    \bibitem[\protect\citeauthoryear{Mathewson, Cleary \& Murray}{1975}]{Mathewson1975}
    Mathewson D.~S., Cleary M.~N., Murray J.~D., 1975, ApJ, 195, L97
    \bibitem[\protect\citeauthoryear{Miller, Bregman \& Wakker}{2009}]{Miller2009}
    Miller E.~D., Bregman J.~N., Wakker B.~P., 2009, ApJ, 692, 470
    \bibitem[\protect\citeauthoryear{Pietrzy\'{n}ski et al.}{2006}]{Pietrzynski2006}
    Pietrzy\'{n}ski G.\ et al., 2006, AJ, 132, 2556
    \bibitem[\protect\citeauthoryear{Puche, Carignan \& Wainscoat}{1991}]{Puche1991}
    Puche D., Carignan C., Wainscoat R.~J., 1991, AJ, 101, 447
    \bibitem[\protect\citeauthoryear{Putman et al.}{2003}]{Putman2003}
    Putman M.~E., Staveley-Smith L., Freeman K.~C., Gibson B.~K., Barnes D.~G., 2003, ApJ, 586, 170
    \bibitem[\protect\citeauthoryear{Putman et al.}{2009}]{Putman2009}
    Putman M.~E.\ et al., 2009, ApJ, 703, 1486
    \bibitem[\protect\citeauthoryear{Robinson \& van~Damme}{1964}]{Robinson1964}
    Robinson B.~J., van~Damme K.~J., 1964, IAUS, 20, 276
    \bibitem[\protect\citeauthoryear{Robinson \& van~Damme}{1966}]{Robinson1966}
    Robinson B.~J., van~Damme K.~J., 1966, AuJPh, 19, 111
    \bibitem[\protect\citeauthoryear{Rodriguez-Fernandez \& Combes}{2008}]{Rodriguez-Fernandez2008}
    Rodriguez-Fernandez N.~J., Combes F., 2008, A\&A, 489, 115
    \bibitem[\protect\citeauthoryear{Rogstad, Lockhart \& Wright}{1974}]{Rogstad1974}
    Rogstad D.~H., Lockhart I.~A., Wright M.~C.~H., 1974, ApJ, 193, 309
    \bibitem[\protect\citeauthoryear{Sancisi \& Allen}{1979}]{Sancisi1979}
    Sancisi R., Allen R.~J., 1979, A\&A, 74, 73
    \bibitem[\protect\citeauthoryear{Schlegel, Barrett \& Singh}{1997}]{Schlegel1997}
    Schlegel E.~M., Barrett P., Singh K.~P., 1997, AJ, 113, 1296
    \bibitem[\protect\citeauthoryear{Seielstad \& Whiteoak}{1965}]{Seielstad1965}
    Seielstad G.~A., Whiteoak J.~B., 1965, ApJ, 142, 616
    \bibitem[\protect\citeauthoryear{Shapiro \& Field}{1976}]{Shapiro1976}
    Shapiro P.~R., Field G.~B., 1976, ApJ, 205, 762
    \bibitem[\protect\citeauthoryear{Sofue \& Rubin}{2001}]{Sofue2001}
    Sofue Y., Rubin V., 2001, ARA\&A, 39, 137
    \bibitem[\protect\citeauthoryear{Spitoni, Recchi \& Matteucci}{2008}]{Spitoni2008}
    Spitoni E., Recchi S., Matteucci F., 2008, A\&A, 484, 743
    \bibitem[\protect\citeauthoryear{Stobbart, Roberts \& Warwick}{2006}]{Stobbart2006}
    Stobbart A.-M., Roberts T.~P., Warwick R.~S., 2006, MNRAS, 370, 25
    \bibitem[\protect\citeauthoryear{Tanaka et al.}{2011}]{Tanaka2011}
    Tanaka M., Chiba M., Komiyama Y., Guhathakurta P., Kalirai J.~S., 2011, ApJ, 738, 150
    \bibitem[\protect\citeauthoryear{Thilker et al.}{2004}]{Thilker2004}
    Thilker D.~A., Braun R., Walterbos R.~A.~M., Corbelli E., Lockman F.~J., Murphy E., Maddalena R., 2004, ApJ, 601, L39
    \bibitem[\protect\citeauthoryear{Thom et al.}{2008}]{Thom2008}
    Thom C., Peek J.~E.~G., Putman M.~E., Heiles C., Peek K.~M.~G., Wilhelm R., 2008, ApJ, 684, 364
    \bibitem[\protect\citeauthoryear{van der Hulst et al.}{1992}]{vanderHulst1992}
    van~der~Hulst J.~M., Terlouw J.~P., Begeman K.~G., Zwitser W., Roelfsema P.~R., 1992, ASP Conf.\ Ser., 25, 131
    \bibitem[\protect\citeauthoryear{van~de~Steene et al.}{2006}]{vandeSteene2006}
    van~de~Steene G.~C., Jacoby G.~H., Praet C., Ciardullo R., Dejonghe H., 2006, A\&A, 455, 891
    \bibitem[\protect\citeauthoryear{Vilardell}{2007}]{Vilardell2007}
    Vilardell F., Jordi C., Ribas I., 2007, A\&A, 473, 847
    \bibitem[\protect\citeauthoryear{Wakker \& van~Woerden}{1997}]{Wakker1997}
    Wakker B.~P., van~Woerden H., 1997, ARA\&A, 35, 217
    \bibitem[\protect\citeauthoryear{Wakker et al.}{2007}]{Wakker2007}
    Wakker B.~P.\ et al., 2007, ApJ, 670, L113
    \bibitem[\protect\citeauthoryear{Westmeier, Braun \& Thilker}{2005}]{Westmeier2005}
    Westmeier T., Braun R., Thilker D.~A., 2005, A\&A, 436, 101
    \bibitem[\protect\citeauthoryear{Westmeier, Br\"{u}ns \& Kerp}{2008}]{Westmeier2008}
    Westmeier T., Br\"{u}ns C., Kerp J., 2008, MNRAS, 390, 1691
    \bibitem[\protect\citeauthoryear{Westmeier, Braun \& Koribalski}{2011}]{Westmeier2011}
    Westmeier T., Braun R., Koribalski B.~S., 2011, MNRAS, 410, 2217
  \end{thebibliography}
\end{document}